\documentclass[a4paper,11pt]{article}
\pdfoutput=1
\usepackage{jcappub} 
\usepackage[latin9]{inputenc}
\usepackage{textcomp}
\usepackage{amsmath}
\usepackage{amssymb}
\usepackage{color}
\usepackage{graphicx}
\usepackage{amsmath}
\usepackage{subfigure}
\usepackage{amssymb}
\usepackage{amsfonts}
\usepackage{mathtools}
\usepackage{amssymb}
\usepackage{enumerate}
\usepackage{xcolor}
\usepackage{bm}
\usepackage{mathrsfs}
\usepackage{epstopdf}
\usepackage{url}
\usepackage{footnote}
\usepackage{textcomp}
\usepackage{dsfont}
\usepackage{ulem}
\usepackage{hyperref}
\usepackage{enumerate}   
\usepackage{appendix}
\usepackage{textcomp}
\usepackage{tipa}

\makeatletter

\newcommand{\be}[1]{\begin{equation}\label{#1}}
\newcommand{\ee}{\end{equation}}
\newcommand{\ba}[1]{\begin{eqnarray}\label{#1}}
\newcommand{\ea}{\end{eqnarray}}

\newcommand*{\rom}[1]{\expandafter\@slowromancap\romannumeral #1@}
\makeatother
\newcommand\ztwo{{$\mathbb{Z}_2$ }}

\makeatother
\begin{document}

\title{Observational features of reflection asymmetric black holes}
\author[a]{Che-Yu Chen}
\affiliation[a]{RIKEN iTHEMS, Wako, Saitama 351-0198, Japan}

\emailAdd{b97202056@gmail.com}

\author[b,c,d]{Hung-Yi Pu}
\affiliation[b]{Department of Physics, National Taiwan Normal University, Taipei 116, Taiwan}
\affiliation[c]{Institute of Astronomy and Astrophysics, Academia Sinica, Taipei 10617, Taiwan}
\affiliation[d]{Physics Division, National Center for Theoretical Sciences, Taipei 10617, Taiwan}
\emailAdd{hypu@gapps.ntnu.edu.tw}

\abstract{The Kerr spacetime is symmetric with respect to a well-defined equatorial plane. When testing the equatorial reflection symmetry of an isolated black hole, one is at the same time testing the Kerr hypothesis in General Relativity. In this work, we investigate the possible observational features when a Keplerian disk is surrounding a rotating black hole without reflection symmetry. When such symmetry is broken, generically, the photon trajectories around the black hole and the Keplerian orbits on the accretion disk are distorted vertically away from the equatorial plane by an amount that depends on their distance to the black hole. In the reflection asymmetric spacetime we are considering, these two kinds of orbits are distorted in opposite directions. Interestingly, while the size and shape of black hole shadows closely resemble those of Kerr black holes, distinct observational characteristics can emerge in the disk image and emission line profiles. When observing the disk edge-on, a pronounced concave shape may appear along its innermost edge on the incoming side. Furthermore, distinctive horn-like features might be observed on the spectral line profile at the blue-shifted side. These special features can serve as compelling indicators of the reflection asymmetry present in rotating black holes.}

\maketitle
\flushbottom

\section{Introduction}

One important prediction about black holes based on General Relativity (GR) is the Kerr hypothesis, which states that an isolated black hole in the universe is described by the Kerr spacetime. It is well known that the Kerr spacetime is stationary and has a well-defined axis of symmetry with respect to which the black hole is spinning. Also, the Kerr black hole satisfies the no-hair theorem in the sense that the whole black hole geometry is characterized by only its mass $M$ and spin $a$. With the future advancements in gravitational wave detection \cite{LIGOScientific:2016aoc} and black hole imaging \cite{EventHorizonTelescope:2019dse,EventHorizonTelescope:2022wkp}, directly testing the Kerr hypothesis via black hole observations would become a promising way of testing GR \cite{Bambi:2011mj,Barack:2018yly}, which may open a novel window to discover new physics that up to now has not been accessible due to observational limitations.

In addition to axisymmetry and stationarity, the Kerr spacetime has a few other inherent symmetries. First, the Kerr spacetime has $\mathbb{Z}_2$ symmetry, i.e., it has a well-defined equatorial plane that stands for a surface of symmetry of the spacetime. Second, the Kerr spacetime has a hidden symmetry that corresponds to the existence of a non-trivial rank-2 Killing tensor. This symmetry is associated with a conserved quantity in the geodesic dynamics, i.e., the Carter constant \cite{Carter:1968rr}, such that the geodesic dynamics is Liouville integrable. The third symmetry is circularity \cite{Papapetrou:1966zz}, which implies the existence of a foliation of two-dimensional surfaces that are everywhere orthogonal to the surfaces of transitivity, i.e., the surfaces spanned by the two Killing vectors associated with the axisymmetry and stationarity \cite{Xie:2021bur}. From the perspective of testing the Kerr hypothesis, the violation of any of the individual symmetries mentioned above for an isolated black hole is a smoking gun for the violation of the hypothesis, which may hint toward some physics beyond GR. Therefore, it becomes crucial to understand what observational features would appear when the individual symmetry is broken. For example, if the black hole has no $\mathbb{Z}_2$ symmetry, the accretion thin disk around the black hole generically acquires a curved surface. This is because the orbits of massive particles moving on the disk would be shifted by an amount that depends on the distance to the black hole \cite{Chen:2021ryb}. Furthermore, the violation of Liouville integrability would lead to chaotic behaviors for orbital dynamics around the black hole, which may leave interesting imprints on the black hole images \cite{Cunha:2015yba,Cunha:2016bjh,Kostaros:2021usv} and on the gravitational waves emitted from small objects orbiting the black hole \cite{Apostolatos:2009vu,Lukes-Gerakopoulos:2010ipp,Destounis:2021mqv,Destounis:2021rko,Destounis:2023khj}. Finally, the violation of circularity deprives the spacetime of the non-trivial rank-2 Killing tensor \cite{Delaporte:2022acp}, which may also lead to chaotic orbital dynamics \cite{Zhou:2021cef,Chen:2023gwm}. In this paper, we will focus on the observational features induced by the violation of $\mathbb{Z}_2$ symmetry of black holes.

The violation of $\mathbb{Z}_2$ symmetry of spacetimes can appear in various scenarios. One possibility is the microstate geometries with horizon-size structures that are inspired by string theory \cite{Bianchi:2020bxa,Bianchi:2020miz,Bah:2021jno}. These compact objects have rich multipole structures such that the $\mathbb{Z}_2$ symmetry may be broken in a model-dependent manner. The detectability of such effects through future space-based gravitational wave observations has been discussed \cite{Fransen:2022jtw}. In addition, the $\mathbb{Z}_2$ asymmetric black holes may appear in the effective field theories of quantum gravity that contain parity-violating terms in the gravitational action. This includes the black hole solutions found in Ref.~\cite{Cardoso:2018ptl} where the action of the theory contains explicit parity-violating curvature invariants. Also, rotating black holes without $\mathbb{Z}_2$ symmetry have been identified in the models of Ref.~\cite{Cano:2019ore,Cano:2022wwo} whose gravitational actions contain a Chern-Simons term and a Gauss-Bonnet term coupled through a dynamical scalar field. Due to the complexity of the equations of motion, the rotating black hole solutions in these parity-violating theories can only be solved perturbatively. Recently, an exact rotating black hole solution in a particular parity-violating theory has been obtained through an invertible conformal transformation \cite{Tahara:2023pyg}. The conformal factor explicitly depends on the Chern-Simons term, so the spacetime breaks the $\mathbb{Z}_2$ symmetry. The viability of this approach is based on the invertibility of the conformal transformation, which ensures the existence of a gravitational theory to which the conformally transformed metric is an exact solution. We would like to emphasize that in the aforementioned $\mathbb{Z}_2$ asymmetric black hole solutions, the parameters that break the $\mathbb{Z}_2$ symmetry simultaneously break the Liouville integrability for geodesic dynamics. Even for the conformal Kerr metric proposed in Ref.~\cite{Tahara:2023pyg}, the geodesic dynamics for massive particles is not integrable despite the integrability for photon geodesic. In addition to the rotating black hole solutions that are subject to specific gravitational theories, one may construct $\mathbb{Z}_2$ asymmetric black hole metrics from a theory-agnostic point of view. In Refs.~\cite{Chen:2020aix,Chen:2021ryb,Chen:2022lct}, a phenomenological model for $\mathbb{Z}_2$ asymmetric black hole spacetimes has been proposed, which can be regarded as a parameterized metric describing reflection asymmetric black holes that still maintain the Liouville integrability for geodesic dynamics.

To investigate the possibility of testing black hole $\mathbb{Z}_2$ symmetry through observations, one should first understand what the unique features the violation of $\mathbb{Z}_2$ symmetry may lead to. Naively, the lack of $\mathbb{Z}_2$ symmetry makes the definition of the standard equatorial plane ambiguous. In fact, the standard equatorial plane, i.e., $\theta=\pi/2$ in the Boyer-Lindquist coordinate system, is not a plane of symmetry anymore in this case. Specifically, a moving particle at $\theta=\pi/2$ with initial velocities tangent to that plane receives extra external off-plane forces induced by the $\mathbb{Z}_2$ asymmetry \cite{Datta:2020axm}. Therefore, for an axisymmetric spacetime that breaks $\mathbb{Z}_2$ symmetry, the circular orbits that are normal to the axis of symmetry may still exist, but they do not lie on the standard equatorial plane. One direct consequence is that if we regard the accretion thin disk as a collection of timelike circular orbits with different radii, the thin disk would have a curved surface \cite{Chen:2021ryb}. It is then interesting to explore what the observational implications of such a curved disk would be. To address this issue, in this paper, we will focus on the phenomenological model for reflection asymmetric black holes proposed in Refs.~\cite{Chen:2020aix,Chen:2021ryb,Chen:2022lct}, hereafter dubbed No$\mathbb{Z}$  black hole model. The No$\mathbb{Z}$ black hole model provides a theory-agnostic parameterization of $\mathbb{Z}_2$ asymmetric black hole spacetimes. Also, the construction of the model is based on the assumption that the No$\mathbb{Z}$ black hole spacetime still preserves the Liouville integrability of geodesic dynamics. Such a parameterization allows us to focus only on the phenomenological effects specifically induced by the $\mathbb{Z}_2$ asymmetry. In Refs.~\cite{Chen:2020aix,Chen:2022lct}, the shadow images cast by the No$\mathbb{Z}$ black hole were studied. However, the investigation only focused on the critical curve on the image plane, i.e., the closed contour on the image plane defined by the impact parameter of the set of unstable spherical photon orbits \cite{Claudel:2000yi}. It was shown that due to the existence of the non-trivial rank-2 Killing tensor and the Liouville integrability, the critical curve is always symmetric with respect to the horizontal axis of the image plane seen by a distant observer \cite{Chen:2020aix} (see also Refs.~\cite{Cunha:2018uzc,Lima:2021las,Staelens:2023jgr}), irrespective of the inclination angle, i.e., the angle between the line of sight and the black hole rotational axis, of the observer. In Ref.~\cite{Staelens:2023jgr}, it was further argued that the effects on the higher-order lensed images induced by the curved disk surface would be unmeasurably small. In fact, the deviation parameter that controls the $\mathbb{Z}_2$ asymmetry in the No$\mathbb{Z}$ black hole model mainly changes the apparent size of the critical curve. Such a feature enables us to place preliminary constraints on the model using the apparent ring size of the Sgr A* images \cite{Chen:2022lct}.

Considering the No$\mathbb{Z}$ black hole spacetime of Refs.~\cite{Chen:2020aix,Chen:2021ryb,Chen:2022lct}, in this paper, we will explore in depth the possible image features induced by the $\mathbb{Z}_2$ asymmetry. Instead of focusing only on the critical curve or higher-order lensed images, we will pay more attention to the image portion that reflects the disk morphology (e.g. the direct emission), taking into account the fact that the thin disk possesses a curved surface. Based on the results, we will construct the redshift images, the flux images, and the line profiles from the curved disk of the No$\mathbb{Z}$ black hole model. Historically, the modelings of relativistic iron K-shell line profiles from a thin disk around a Kerr black hole have been widely considered \cite{Fabian:1989ej,Laor:1991nc,Bromley:1996wb,fa97}. In addition to the rotation of the accretion disk, the Doppler shifts induced by the frame-dragging effect of a rotating black hole naturally modify the line profiles emerging from the disk, providing helpful diagnostics to interpret the spacetime features, including the black hole spin. Apparently, the line profiles would also be affected by the disk properties \cite{Fuerst:2007am,Wu:2007bq}. Therefore, motivated by how the thin disk around a \ztwo asymmetric black hole deviates from that of a Kerr black hole, here we also aim to explore the observational features of the line emission from the curved disk around a \ztwo asymmetric black hole, and compare with the line emission features of the Kerr black holes.

The paper is organized as follows. In sec.~\ref{sec:th}, we first introduce the phenomenological model for No$\mathbb{Z}$ black holes that we are going to consider in this work. Then, in sec.~\ref{subsec.diskorbit}, we elucidate how the geodesic dynamics is altered in this model as compared with that in the Kerr spacetime. In sec.~\ref{sec:num}, we present the numerical results of our analysis, including the critical curve of the shadow image, the redshifts of the disk morphology, the line emissions, and the spectral profiles. Finally, we conclude in sec.~\ref{sec:con}. The appendix \ref{appeid.a} briefly reviews the construction of our phenomenological black hole model, which was originally proposed in Refs.~\cite{Chen:2020aix,Chen:2021ryb,Chen:2022lct}.

\section{The No$\mathbb{Z}$ black hole metric}\label{sec:th}

The main theme of this paper is to look for the observational features of a black hole spacetime in which the $\mathbb{Z}_2$ symmetry is broken. To focus on the effects from the $\mathbb{Z}_2$ asymmetry, we shall consider a model in which all the symmetries possessed by the Kerr spacetime, except for $\mathbb{Z}_2$, are still preserved. Therefore, we consider the phenomenological model of the No$\mathbb{Z}$ black hole spacetime proposed in Refs.~\cite{Chen:2020aix,Chen:2021ryb,Chen:2022lct}, whose metric $g_{\mu\nu}$ can be expressed as
\begin{align}
g_{tt}=&-1+\frac{2Mr\left(r^2+a^2y^2\right)}{\left(r^2+a^2y^2\right)^2+\left(r^2-2Mr+a^2y^2\right)\tilde\epsilon(y)}\,,\label{gtt}\\
g_{\varphi\varphi}=&\,\left(1-y^2\right)\left(r^2+a^2y^2+\tilde\epsilon(y)\right)\times\nonumber\\&\frac{\left[r^4+a^4y^2+r^2\left(a^2+a^2y^2+\tilde\epsilon(y)\right)+a^2\tilde\epsilon(y)+2Mr\left(a^2-a^2y^2-\tilde\epsilon(y)\right)\right]}{\left(r^2+a^2y^2\right)^2+\left(r^2-2Mr+a^2y^2\right)\tilde\epsilon(y)}\,,\label{gpp}\\
g_{t\varphi}=&-\frac{2Mra\left(1-y^2\right)\left(r^2+a^2y^2+\tilde\epsilon(y)\right)}{\left(r^2+a^2y^2\right)^2+\left(r^2-2Mr+a^2y^2\right)\tilde\epsilon(y)}\,,\label{gtp}\\
g_{rr}=&\,\frac{r^2+a^2y^2+\tilde\epsilon(y)}{r^2-2Mr+a^2}\,,\qquad g_{yy}=\frac{r^2+a^2y^2+\tilde\epsilon(y)}{1-y^2}\label{g23}\,,
\end{align}
where $M$ and $a$ denote the mass and the spin of the black hole, respectively. We choose the Boyer-Lindquist coordinate system $(t,r,y,\varphi)$ with $y\equiv\cos\theta$. The construction of the metric is detailed in Appendix \ref{appeid.a}. The metric functions have an additional deviation function $\tilde\epsilon(y)$ that quantifies the difference of the spacetime as compared with the Kerr one. If $\tilde\epsilon(y)=0$, the spacetime reduces to the Kerr spacetime.

Apparently, the No$\mathbb{Z}$ black hole spacetime is stationary and axisymmetric because there is no explicit $t$ and $\varphi$ dependence in the metric functions. The spacetime is asymptotically flat. In addition, even in the presence of the deviation function $\tilde\epsilon(y)$, the Liouville integrability of the geodesic dynamics is still preserved because there exists a non-trivial rank-2 Killing tensor. The corresponding conserved quantity, i.e., the Carter constant, can be identified, which can be used to recast the geodesic equations into their first-order forms. Besides, in the Boyer-Lindquist coordinate, the event horizon of the No$\mathbb{Z}$ black hole has a constant radius, whose value is determined by the root of the equation $r^2-2Mr+a^2=0$ and is the same as the Kerr one. Essentially, the $\mathbb{Z}_2$ symmetry of the spacetime would be broken as long as the deviation function $\tilde\epsilon(y)$ is not symmetric with respect to $y=0$. As suggested in Appendix \ref{appeid.a}, we will consider $\tilde\epsilon(y)=\epsilon May$ with $\epsilon$ a dimensionless constant for the rest of the paper. Also, we will assume $|\epsilon|\le2$ and $|a|/M<1$ to avoid the possible existence of naked singularities in the spacetime (see Figure 3 of \cite{Chen:2021ryb}). From now on, we will adopt the geometrized unit and use the dimensionless parameter $a_*\equiv a/M$ to quantify the spin value of the black hole.

\subsection{Curved disk and spherical photon orbits}\label{subsec.diskorbit}

The $\mathbb{Z}_2$ asymmetry of a black hole spacetime may affect the motion of particles orbiting the black hole such that a single trajectory behaves differently when it propagates in the north and the south hemispheres. In particular, bound trajectories that are cooped up only in one hemisphere can exist. A direct consequence of the existence of these orbits is that the thin accretion disk around the black hole acquires a curved surface \cite{Chen:2021ryb}. This is because the circular orbits of massive particles that constitute the accretion disk would not be at the standard equatorial plane $y=0$ in general. In addition, the spatial distribution of spherical photon orbits would be deformed such that the circular photon orbits are not at the $y=0$ plane either. The combination of the off-plane behaviors of the accretion disk and photon orbits would give rise to novel features in the disk morphology, black hole images, and line profiles, as we will demonstrate in sec.~\ref{sec:num}. In the rest of this section, we briefly summarize how the structures of accretion thin disks and the spherical photon orbits are altered by the $\mathbb{Z}_2$ asymmetry of the No$\mathbb{Z}$ black hole.

To demonstrate the off-plane behaviors of particle trajectories, we consider the Lagrangian of the geodesic motions:
\begin{equation}
\mathcal{L}=\frac{1}{2}g_{\mu\nu}\dot{x}^\mu\dot{x}^\nu\,,\label{lagrangian}
\end{equation}
where $x^\mu=(t,r,y,\varphi)$ and the dot denotes the derivative with respect to the proper time or the affine parameter for massive particles and photons, respectively. Because the spacetime is stationary and axisymmetric, i.e., the metric functions \eqref{gtt}-\eqref{g23} do not have explicit $t$ and $\varphi$ dependencies, one can define two constants of motion $E\equiv-p_t$ and $L_z\equiv p_\varphi$, where $p_\mu\equiv\partial\mathcal{L}/\partial \dot{x}^\mu$ represent the conjugate momenta of the coordinates $x^\mu$. The two constants of motion $E$ and $L_z$ correspond to the energy and the azimuthal angular momentum of particles at spatial infinity. One can invert the definitions of these constants of motion to get
\begin{align}
\dot{t}&=p^t=\frac{Eg_{\varphi\varphi}+L_zg_{t\varphi}}{g_{t\varphi}^2-g_{tt}g_{\varphi\varphi}}\,,\label{tdot}\\
\dot{\varphi}&=p^\varphi=-\frac{Eg_{t\varphi}+L_zg_{tt}}{g_{t\varphi}^2-g_{tt}g_{\varphi\varphi}}\,,\label{phidot}
\end{align}
where $p^\mu=g^{\mu\nu}p_\nu$.

The Hamiltonian of the Lagrangian \eqref{lagrangian} reads
\begin{equation}
\mathcal{H}=\frac{1}{2}p_\mu p^\mu\,,
\end{equation}
from which the full set of eight geodesic equations can be obtained: $\dot{x}^\mu=\partial\mathcal{H}/\partial p_\mu$ and $\dot{p}_\mu=-\partial\mathcal{H}/\partial x^\mu$. Apparently, the $t$ and $\varphi$ components of these equations are precisely given by Eq.~\eqref{tdot}, Eq.~\eqref{phidot}, and the conservation of $E$ and $L_z$. The other four equations consist of the definitions of two conjugate momenta, i.e., $\dot{r}=p^r$ and $\dot{y}=p^y$, as well as two differential equations that govern the evolution of $p^r$ and $p^y$. These evolution equations are subject to the 4-velocity constraint, which can be written as $\mathcal{H}=-\delta/2$, with $\delta=0$ and $\delta=1$ for photons and massive particles, respectively. The 4-velocity constraint is nothing but the conservation of the rest mass of the particles.

In the most general cases, the right-hand side of the evolution equations for $p^r$ or $p^y$, i.e., $-\partial\mathcal{H}/\partial r$ or $-\partial\mathcal{H}/\partial y$, has explicit dependencies on both $r$ and $y$. Therefore, they can be solved by only using numerical treatments. However, since there exists a non-trivial rank-2 Killing tensor for the No$\mathbb{Z}$ black hole spacetime, the radial and the angular sectors of the evolution equations are separable and can be recast into their first-order forms:
\begin{align}
&\left[r^2+a^2y^2+\epsilon May\right]^2(p^r)^2=\mathcal{R}(r)\,,\label{eqr}\\
&\left[r^2+a^2y^2+\epsilon May\right]^2(p^y)^2=\mathcal{Y}(y)\,,\label{eqy}
\end{align}
where
\begin{align}
\mathcal{R}(r)\equiv&\,\left[E\left(r^2+a^2\right)-aL_z\right]^2-\left(\mathcal{K}+r^2\delta\right)\Delta-\left(L_z-aE\right)^2\Delta\,,\label{rpotential}\\
\mathcal{Y}(y)\equiv&\,\left[\mathcal{K}+\left(L_z-aE\right)^2-a^2y^2\delta-\epsilon May\left(\delta-E^2\right)\right]\left(1-y^2\right)-\left[a\left(1-y^2\right)E-L_z\right]^2\,,\label{ypotential}
\end{align}
$\Delta=r^2-2Mr+a^2$, and $\mathcal{K}$ is a separation constant. On top of $E$, $L_z$, and $\delta$, the separation constant $\mathcal{K}$ is the fourth constant of motion which can also be identified as the Carter constant. Note that according to Eqs.~\eqref{eqy} and \eqref{ypotential}, for orbits that have latitudinal turning points ($p^y=0$) at $y=0$, their constant of motion $\mathcal{K}$ is zero.

\subsubsection{Curved accretion disk}

\begin{figure}[t]
  \centering
 \includegraphics[scale=0.6]{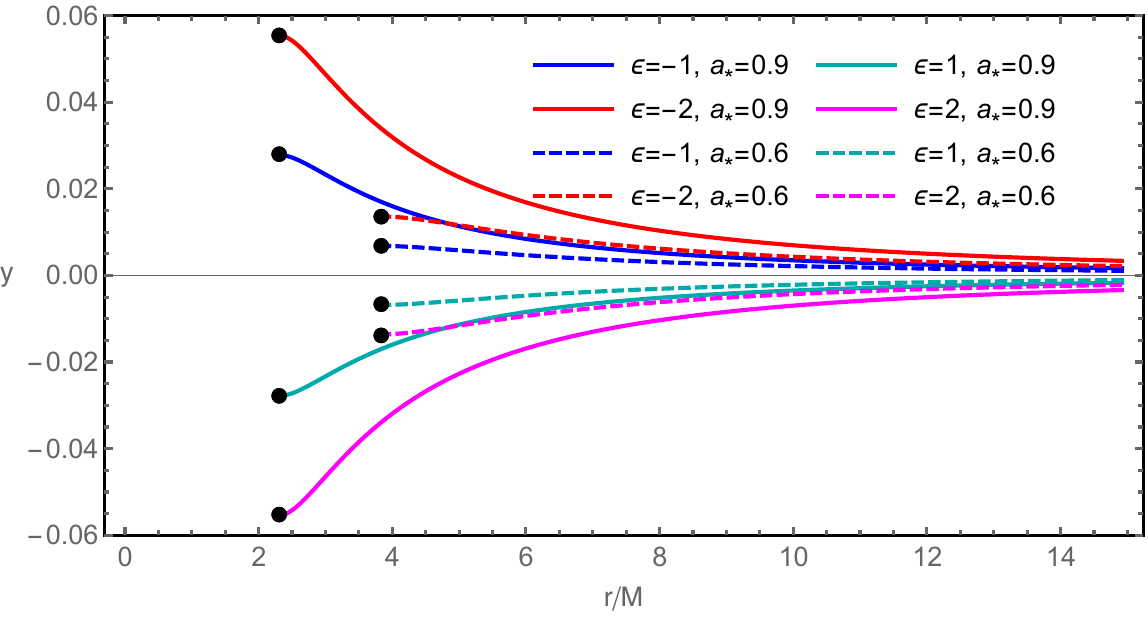}
 \includegraphics[scale=0.6]{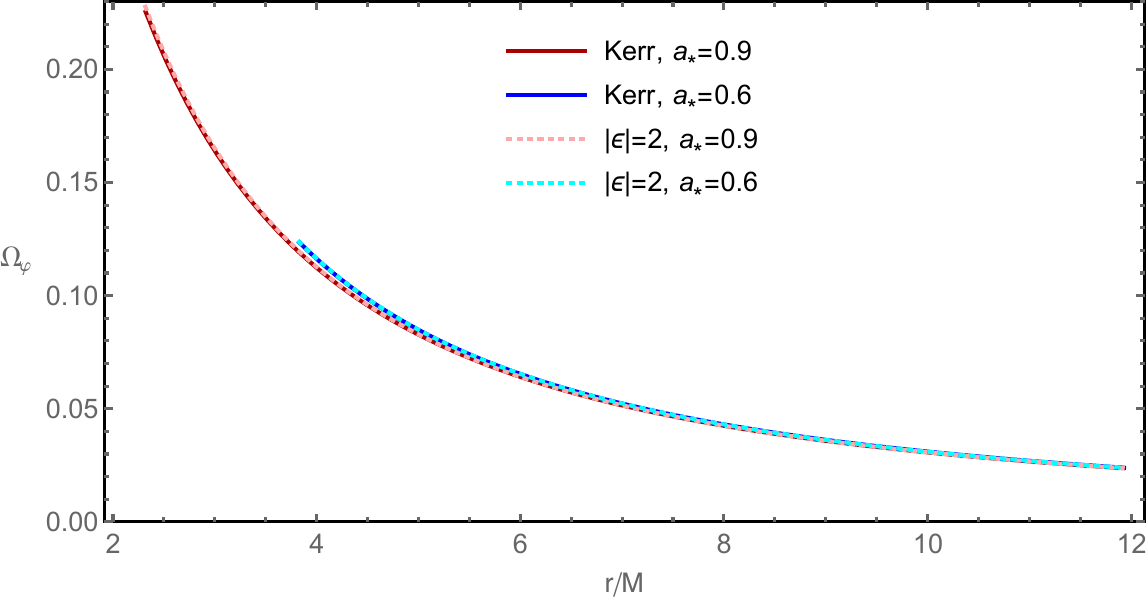}
\\
\caption{Upper: The latitude $y$ with respect to the radius $r$ of circular orbits for massive particles. Each curve terminates at its left edge (black points) which corresponds to the position of the innermost stable circular orbits (ISCO). Lower: The prograde orbital angular velocity $\Omega_\varphi$ of circular orbits with respect to the radius.}
\label{fig:ryplot} 
\end{figure} 

In a sufficiently short timescale during which the radial direction of matter accretion can be neglected, a typical thin accretion disk can be regarded as a collection of circular orbits with different radii of massive particles. To demonstrate that the thin accretion disk acquires a curved surface, which was originally pointed out in Ref.~\cite{Chen:2021ryb}, we consider timelike geodesics, i.e., $\delta=1$. Then, we use the following expression for geodesic equations:
\begin{equation}
\frac{d}{d\tau}\left(g_{\mu\nu}\dot{x}^\nu\right)=\frac{1}{2}\left(\partial_\mu g_{\nu\rho}\right)\dot{x}^\nu\dot{x}^\rho\,,\label{geodeeqalt}
\end{equation}
where $\tau$ denotes the proper time of particles and the dot here denotes the derivative with respect to $\tau$. Assuming that the rotation axis of accreting matters is aligned with that of the black hole spin, the conditions for each circular orbit are  $\dot{r}=\ddot{r}=\dot{y}=\ddot{y}=0$. Therefore, the $r$ and $y$ components of Eq.~\eqref{geodeeqalt} can be written as
\begin{align}
\partial_rg_{tt}+2\left(\partial_rg_{t\varphi}\right)\Omega_\varphi+\left(\partial_rg_{\varphi\varphi}\right)\Omega_\varphi^2=0\,,\label{convenr}\\
\partial_yg_{tt}+2\left(\partial_yg_{t\varphi}\right)\Omega_\varphi+\left(\partial_yg_{\varphi\varphi}\right)\Omega_\varphi^2=0\,,\label{conveny}
\end{align}
where $\Omega_\varphi\equiv\dot\varphi/\dot{t}$ is the orbital angular velocity of the circular orbits. Using Eq.~\eqref{convenr}, the orbital angular velocity can be expressed as
\begin{equation}
\Omega_\varphi=\frac{-\partial_rg_{t\varphi}\pm\sqrt{\left(\partial_rg_{t\varphi}\right)^2-\left(\partial_rg_{tt}\right)\left(\partial_rg_{\varphi\varphi}\right)}}{\partial_rg_{\varphi\varphi}}\,,\label{phidtd}
\end{equation}
where the plus and minus branches correspond to prograde and retrograde orbits, respectively. Given a set of values of $\epsilon$ and $a_*\equiv a/M$, and then inserting the expression \eqref{phidtd} into Eq.~\eqref{conveny}, one can solve Eq.~\eqref{conveny} to get the latitude $y$ associated with each circular orbit of radius $r$. The results are shown in the upper panel of Fig.~\ref{fig:ryplot}. In this figure, we choose some values of $a_*$ and $\epsilon$, then show how the circular orbits with different radii are shifted away from the usual equatorial plane $y=0$ (see also Figure 2 of Ref.~\cite{Chen:2021ryb}). One can see that for a given set of $a_*\ne0$ and $\epsilon\ne0$, each circular orbit lies on its own plane that is parallel to but shifted away from the usual equatorial plane. Due to the asymptotic flatness, the circular orbit with a larger radius lies on a plane closer to the standard equatorial plane, i.e., it has a smaller $|y|$. Therefore, the thin accretion disk, which consists of a collection of circular orbits with different radii, has a curved surface. In addition, the orbits with positive (negative) $\epsilon$ have a negative (positive) value of $y$. Note that the deviation from the equatorial plane is quite limited, with $|\Delta y|\le 0.06$. In the lower panel of Fig.~\ref{fig:ryplot}, we show the prograde orbital angular velocity $\Omega_\varphi$ with respect to the radius of the orbits. One sees that the angular velocity depends more on the values of the black hole spin $a_*$ than on $\epsilon$. For a fixed spin value, the curves of orbital angular velocity with different $\epsilon$ almost overlap. Therefore, when $\epsilon$ varies, the off-plane distribution of the disk has more significant effects on the disk images, as compared with the contributions from the changes of $\Omega_\varphi$.

Note that in Ref.~\cite{Chen:2021ryb}, the relation between the latitude $y$ and radius $r$ of each circular orbit is obtained by first parameterizing $E$ and $L_z$ by using the radius and the Carter constant of the orbit \cite{Teo:2020sey}, then solving $\mathcal{Y}=d\mathcal{Y}/dy=0$ to get the corresponding $y$ and $\mathcal{K}$ simultaneously. That method requires the existence of the Carter constant such that the geodesic equations are separable as those in Eqs.~\eqref{eqr} and \eqref{eqy}. In addition, it requires the corresponding radial potential $\mathcal{R}(r)$ to be exactly given by Eq.~\eqref{rpotential}, otherwise the parameterization provided in Ref.~\cite{Teo:2020sey} is not valid anymore. On the other hand, the method used in the present paper applies to general scenarios such as the spacetimes that do not have integrable orbital dynamics. Most importantly, we confirm that these two methods agree perfectly.

\subsubsection{Spherical photon orbits}

\begin{figure*}[t]
  \centering
 \includegraphics[scale=0.36]{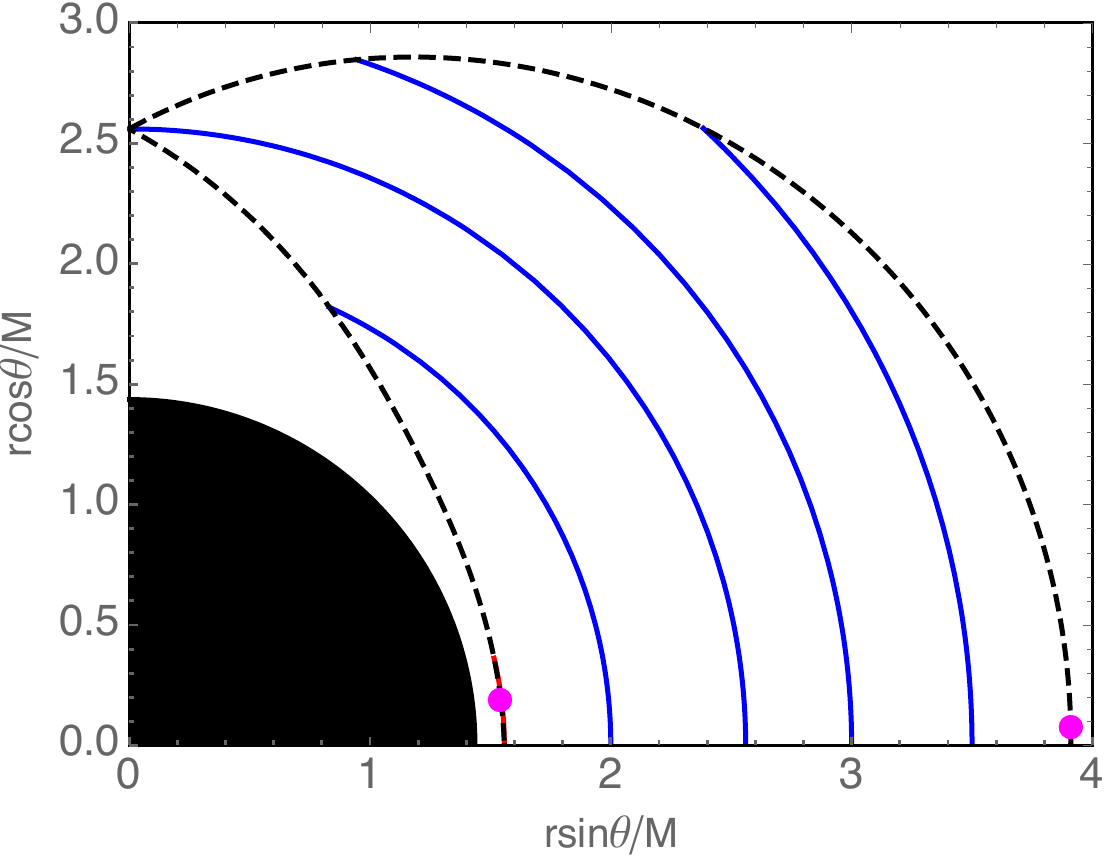}
 \includegraphics[scale=0.36]{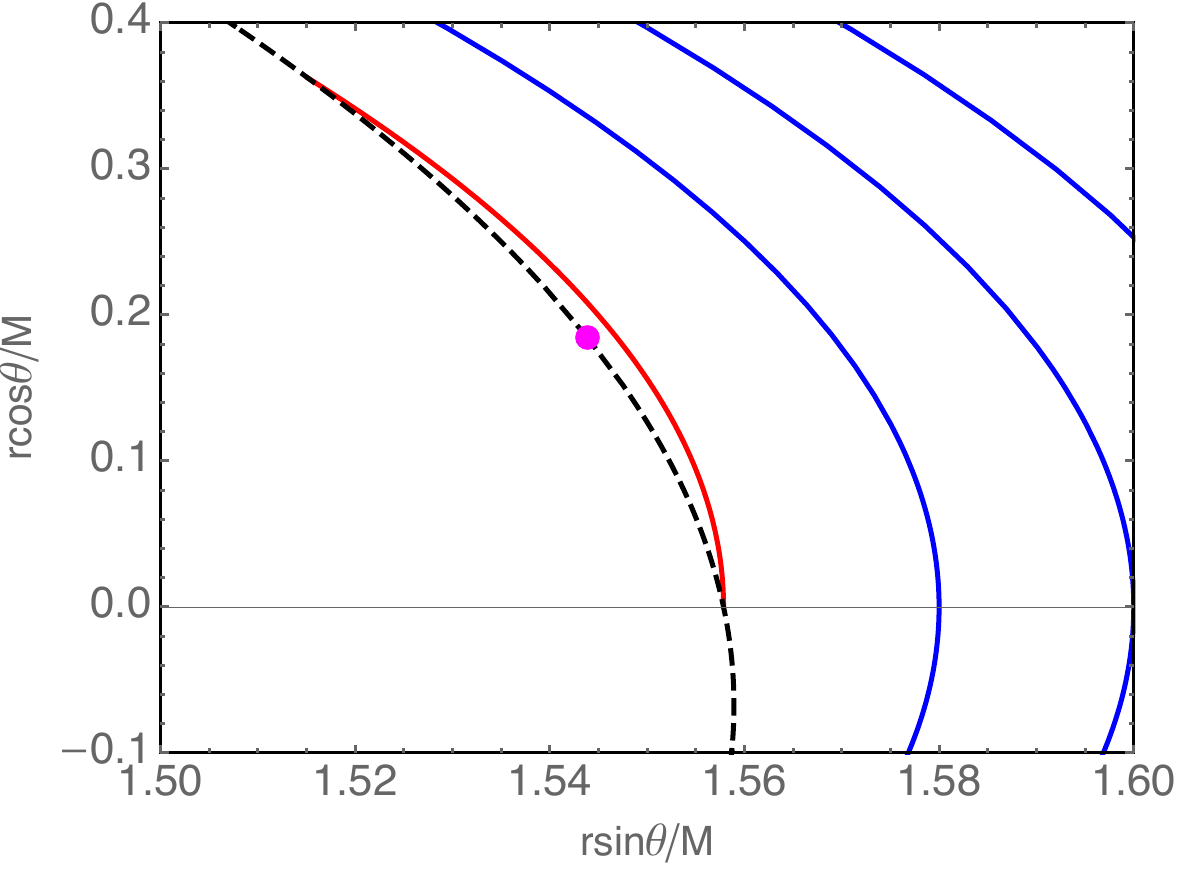}\\
\caption{Left: The structure of spherical photon orbits of No$\mathbb{Z}$ black holes with $a_*=0.9$ and $\epsilon=2$. Blue curves represent spherical orbits with different radii. The orbits have latitudinal turning points on the dashed curves which are defined by $\mathcal{Y}(y)=0$. The magenta points represent the prograde and retrograde circular orbits. Right: The orbits near the prograde circular one. The red curve represents the spherical orbit that has one latitudinal turning point at $y=0$.}
\label{fig:psstructure} 
\end{figure*} 

Let us now discuss the photon geodesic equations ($\delta=0$) and the set of spherical photon orbits around the black hole. For Kerr black holes, each spherical photon orbit has its own radius $r$ and the orbit oscillates in its latitudinal direction. The radii of these orbits are bounded by those of the prograde and retrograde circular photon orbits (or light rings \cite{Cunha:2017eoe}), i.e., $r_-\le r\le r_+$, with
\begin{equation}
r_\mp=2M\left[1+\cos\left(\frac{2}{3}\cos^{-1}\left(\mp|a_*|\right)\right)\right]\,.
\end{equation}
where $r_-$ and $r_+$ denote the radii of the prograde and retrograde circular photon orbits, respectively. Because of the equatorial reflection symmetry of the Kerr spacetime, these two orbits are completely at the equatorial plane $y=0$. However, when the $\mathbb{Z}_2$ symmetry is broken, the circular photon orbits may still exist but would be shifted away from $y=0$. We consider the No$\mathbb{Z}$ black hole spacetime and identify the locations of these two circular photon orbits. These two orbits satisfy $\dot{y}=\ddot{y}=\dot{r}=\ddot{r}=0$, where the dot here stands for the derivative with respect to the affine parameter $\lambda$. Therefore, Eqs.~\eqref{convenr} and \eqref{conveny} are still applicable. To identify the circular orbits, we insert Eq.~\eqref{phidtd}, which is obtained from Eq.~\eqref{convenr}, into Eq.~\eqref{conveny}. The resulting equation and the 4-velocity constraint $\mathcal{H}=0$ form a set of coupled equations that depend only on $r$ and $y$. Solving these equations, one can get the Boyer-Lindquist radius $r$ and the latitude $y$ of the prograde and the retrograde circular photon orbits. For $a_*=0.9$ and $\epsilon=2$, the prograde circular photon orbit has $r=1.55486M$ and $y=0.119037$. The retrograde orbit has $r=3.91052M$ and $y=0.0195954$. See the magenta points in Fig.~\ref{fig:psstructure}.    

In Fig.~\ref{fig:psstructure}, we demonstrate how the spherical photon orbits distribute around the No$\mathbb{Z}$ black hole. Here we choose $a_*=0.9$ and $\epsilon=2$. The shaded region is the black hole interior. Different spherical photon orbits have different radii, which are represented by the blue and red curves. Each orbit has its latitudinal turning points that are determined by $\mathcal{Y}(y)=0$. The collection of the latitudinal turning points of all spherical photon orbits forms a boundary of these orbits, i.e., the dashed curves in Fig.~\ref{fig:psstructure}. For black holes that respect $\mathbb{Z}_2$ symmetry, the dashed curves are symmetric with respect to $y=0$ (see Fig. 21 of Ref.~\cite{Perlick:2004tq} for the structure of spherical photon orbits around Kerr black holes). However, generically, when the reflection symmetry is broken, the latitudinal turning points of a spherical photon orbit are not symmetric with respect to $y=0$. As can be seen from the magenta points, the two circular orbits are not at the equatorial plane $y=0$. In particular, Some spherical photon orbits can have motion completely within one hemisphere because their latitudinal turning points are both on one side of the $y=0$ plane. This can be better understood from the right panel of Fig.~\ref{fig:psstructure}, in which we zoom in on the region near the magenta point that represents the prograde circular orbit. Near the prograde circular orbit, we show a spherical photon orbit with a radius slightly larger than that of the circular one, and it has one latitudinal turning point at $y=0$ (the red curve in Fig.~\ref{fig:psstructure}). This orbit has $\mathcal{K}=0$ as can be shown straightforwardly using Eqs.~\eqref{eqy} and \eqref{ypotential}. The spherical photon orbits with radii within the range between that of the red curve and the magenta point are completely within one hemisphere. This is a manifestation of $\mathbb{Z}_2$ asymmetry of the spacetime.

The fact that the circular orbits of both massive and massless particles are away from the $y=0$ plane is expected to be a common feature when the $\mathbb{Z}_2$ symmetry is broken.{\footnote{One exception is the model proposed in Ref.~\cite{Tahara:2023pyg}. In this model, the photon geodesic dynamics is completely identical to that of the Kerr spacetime because of the conformal structure.}} One interesting feature of the model considered here is that the circular orbits of massive particles and photons are located in opposite hemispheres. For $\epsilon>0$, the circular photon orbits have $y>0$, but the orbits for massive particles as well as the thin disk structure are shifted to the hemisphere with $y<0$.

\section{\label{sec:num}Numerical results}
In this section, we numerically calculate the geodesics in the No$\mathbb{Z}$ black hole spacetime described by the metric functions in Eqs.~\eqref{gtt}-\eqref{g23}. The null geodesics are numerically computed by solving the set of eight equations which consists of ($p^{\mu}=\dot{x}^\mu, \dot{p}^\mu$), where $x^\mu=(t,r,y,\varphi)$ and the dot denotes the derivative with respect to the affine parameter $\lambda$. The geodesics are solved by the Runge-Kutta method, performed by a modified version of the GPU-based radiative-transfer code\footnote{\href{https://github.com/hungyipu/Odyssey}{https://github.com/hungyipu/Odyssey}} {\it Odyssey} \cite{Pu:2016eml}. While there are conserved quantities along geodesics,\footnote{The Carter constant in the No$\mathbb{Z}$ black hole spacetime considered here can be defined in the standard way by taking $K_{\mu\nu}p^\mu p^\nu$, where $K_{\mu\nu}$ is the non-trivial rank-2 Killing tensor.} we do not need the information of them when solving the geodesics in this numerical scheme. Instead, these conserved quantities can serve as a constraint, and one can monitor if these values remain constant within acceptable errors along geodesics in the numerical computation. A similar numerical approach has also been applied to explore the images of black holes dressed with supertranslation hairs \cite{Lin:2022ksb}. In the present work, we consider the No$\mathbb{Z}$ black hole spacetime with $|\epsilon|\le2$ and $|a_*|<1$, as in this choice of the parameter range it is guaranteed that no naked singularity would appear outside the event horizon \cite{Chen:2021ryb}. Also, we will only focus on the prograde orbits for the Keplerian disk.

\subsection{Black Hole Shadow}\label{sec:bh_shadow}

\begin{figure*}
  \centering
 \includegraphics[scale=0.4]{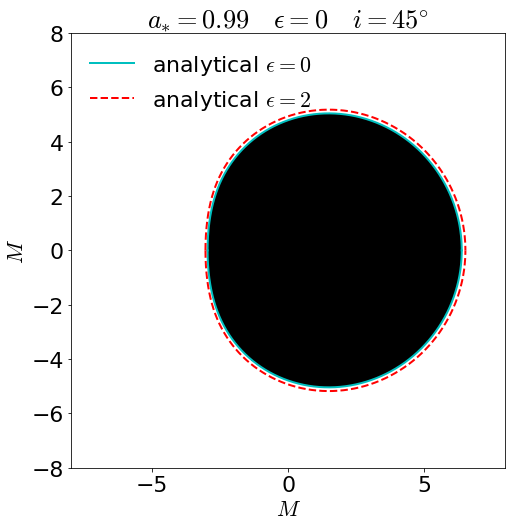}
 \includegraphics[scale=0.4]{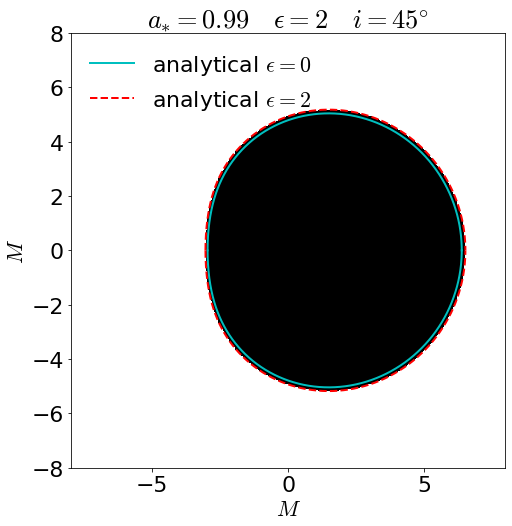}\\
 \includegraphics[scale=0.4]{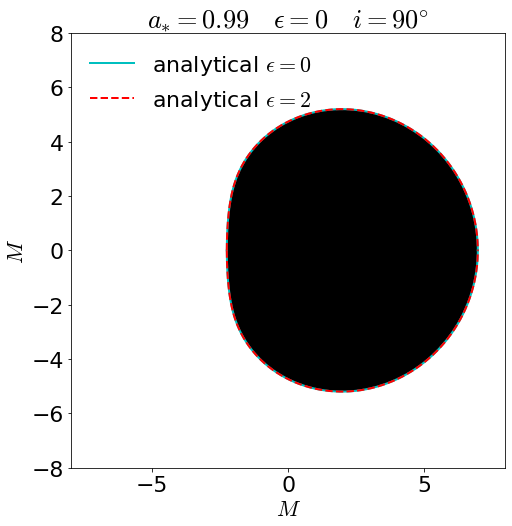}
 \includegraphics[scale=0.4]{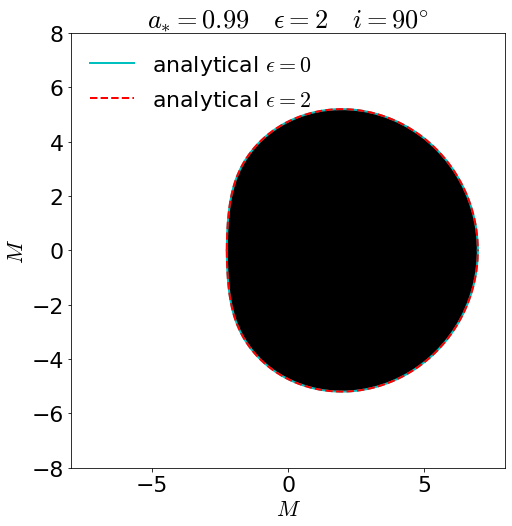}
\caption{Numerically computed black hole shadow images at different viewing angles with respect to the black hole rotational axis, $i$. The dimensionless black hole spin is $a_{*}=0.99$. 
The cases for Kerr ($\epsilon=0$) and the No$\mathbb{Z}$ black holes are shown in the left and right panels, respectively. The image resolution is 512  $\times$ 512 pixels. The black-shaded regions in all plots have boundaries that are consistent with the analytical results given in Refs.~\cite{Chen:2020aix,Chen:2022lct} (the cyan and red curves). When $i=90^{\circ}$, the shapes of the black hole shadows are the same for Kerr and the No$\mathbb{Z}$ black holes, as already pointed out in Refs.~\cite{Chen:2020aix,Chen:2022lct}.}
\label{fig:shadow} 
\end{figure*} 

As a validation test for our numerical code, we first compare the black hole shadows \cite{Falcke:1999pj}, or more precisely, the critical curve \cite{Bardeen:1973} on the image plane that corresponds to the impact parameters of the spherical photon orbits, computed by using numerical and analytic methods. The analytic method was performed in Refs.~\cite{Chen:2020aix,Chen:2022lct}, and its validity is based on the existence of the non-trivial rank-2 Killing tensor as well as the separability of the geodesic equations. Then, we compare the shadows obtained by these two methods to check the consistency between our numerically computed black hole shadow and the analytical results. In Fig.~\ref{fig:shadow}, we present a comparison between the Kerr black hole shadows ($\epsilon=0$) and those of the No$\mathbb{Z}$ black hole ($\epsilon=2$). We fix the dimensionless spin parameter $a_*=0.99$. The shadows are depicted at various observer viewing angles $i$ relative to the black hole's rotational axis, specifically, $i=(45^{\circ}, 90^{\circ})$. In the numerical method, the morphology of these black hole shadows is obtained by numerically tracing the null geodesics from the image plane of a distant observer (assumed to be at $r=10^{3} M$) backward in time, and checking if the geodesics are captured by the black hole. Our analytical results, on the other hand, are calculated with the assumption that the observer is at spatial infinity. According to Fig.~\ref{fig:shadow}, we confirm that the shadows calculated by our numerical method (the boundaries of the black-shaded regions) agree quite well with the ones obtained by the analytic methods (the red and cyan contours).

\begin{figure*}
  \centering
\includegraphics[width=0.95\textwidth]{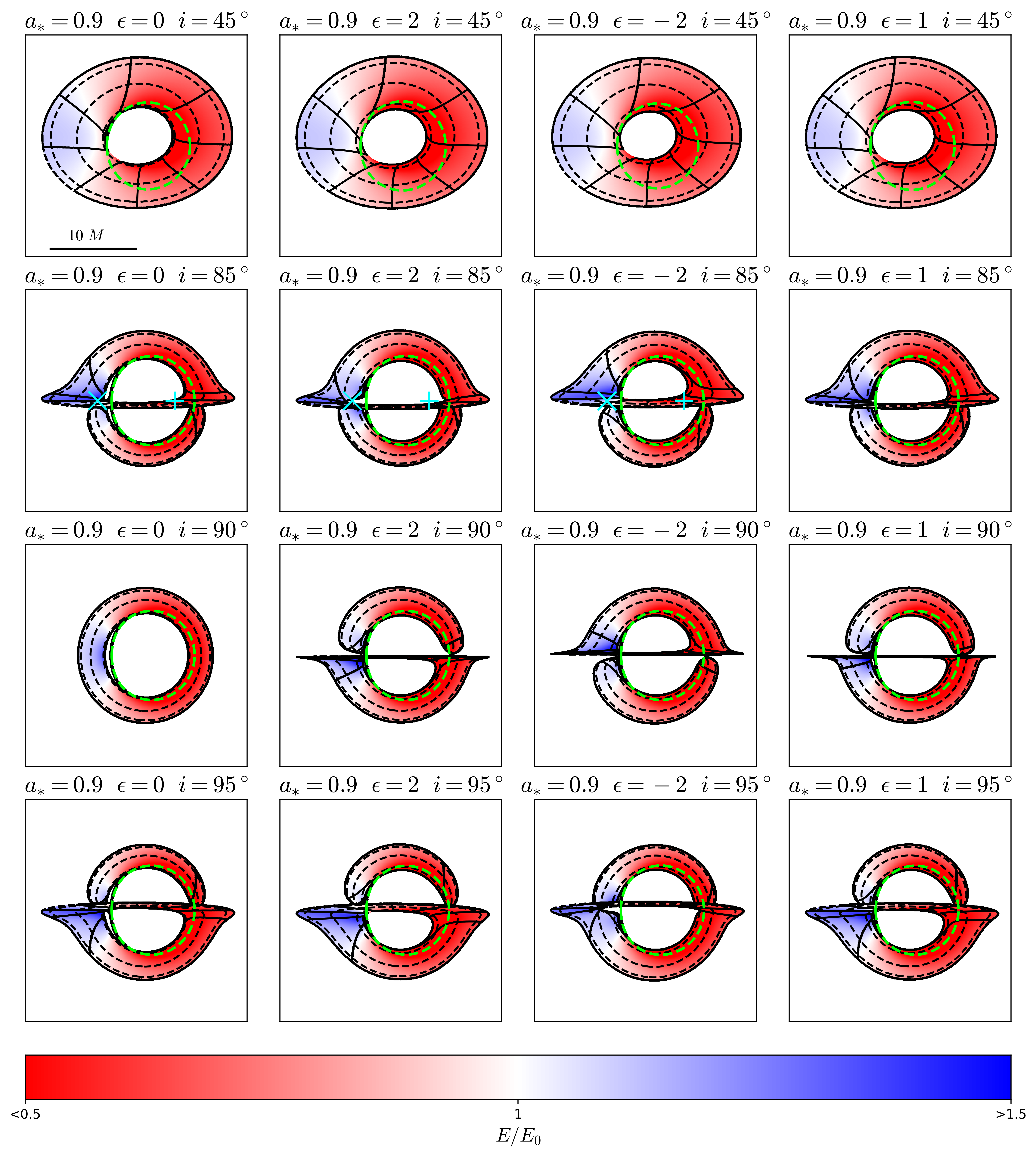}\\
\caption{Energy shifts $E/E_0$ of a thin Keplerian disk consisting of circular orbits around Kerr black holes ($\epsilon=0$) and the No$\mathbb{Z}$ black holes ($\epsilon= (1,~\pm2)$), with $a_{*}=0.9$ and various viewing angles $i$.  The blue-shifted ($E/E_{0}>1$) and red-shifted ($E/E_{0}<1$) parts are respectively indicated by blue and red colors (see Eq.~\eqref{eq:redshift}). The projected black hole spin axis is pointing upward in all the plots.  The outer edge of the disk is assumed to be $r_{\rm out}=10 M$, and the inner edge of the disk is located at the ISCOs, whose radius is a function of $a_*$ and $\epsilon$. Contours of constant $r/M=(3,6,9)$ and $\varphi=n\pi/4$, with $n=(0,1,\ldots7)$ on the disk surface are overlapped. The cyan `$\times$' and `$+$' symbols in the plots with $i=85^{\circ}$ indicate the light rays that reach the observer at the points $(\mp 4.5, 0)$ on the image plane. The boundaries of the black hole shadows are shown by the green dashed curves.}
\label{fig:disk_z} 
\end{figure*}

\begin{figure}
  \centering
 \includegraphics[trim={0 1cm 0 0cm},clip,width=0.65\textwidth]{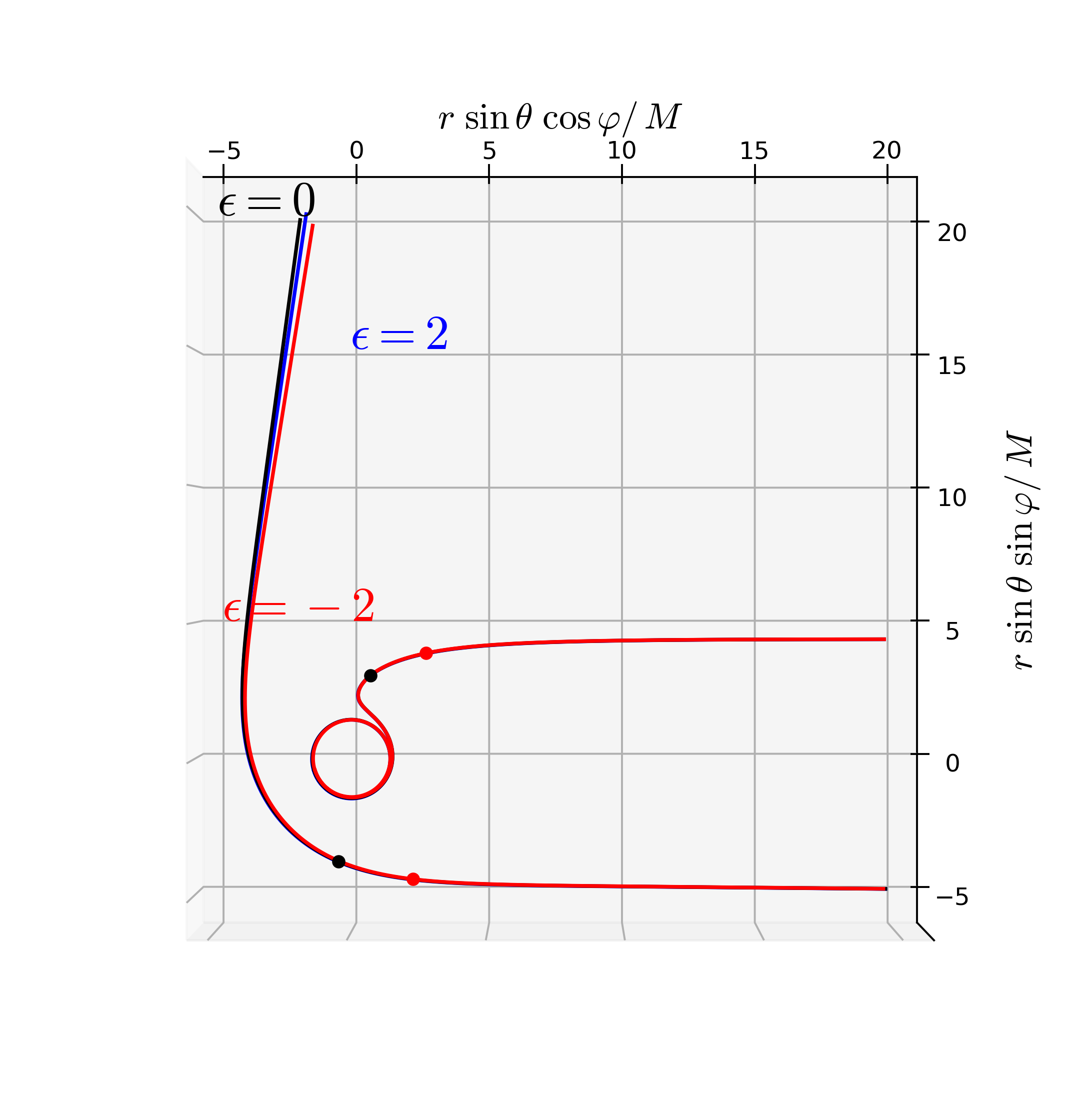}\\
  \includegraphics[trim={0 4cm 0 5cm},clip,width=0.65\textwidth]{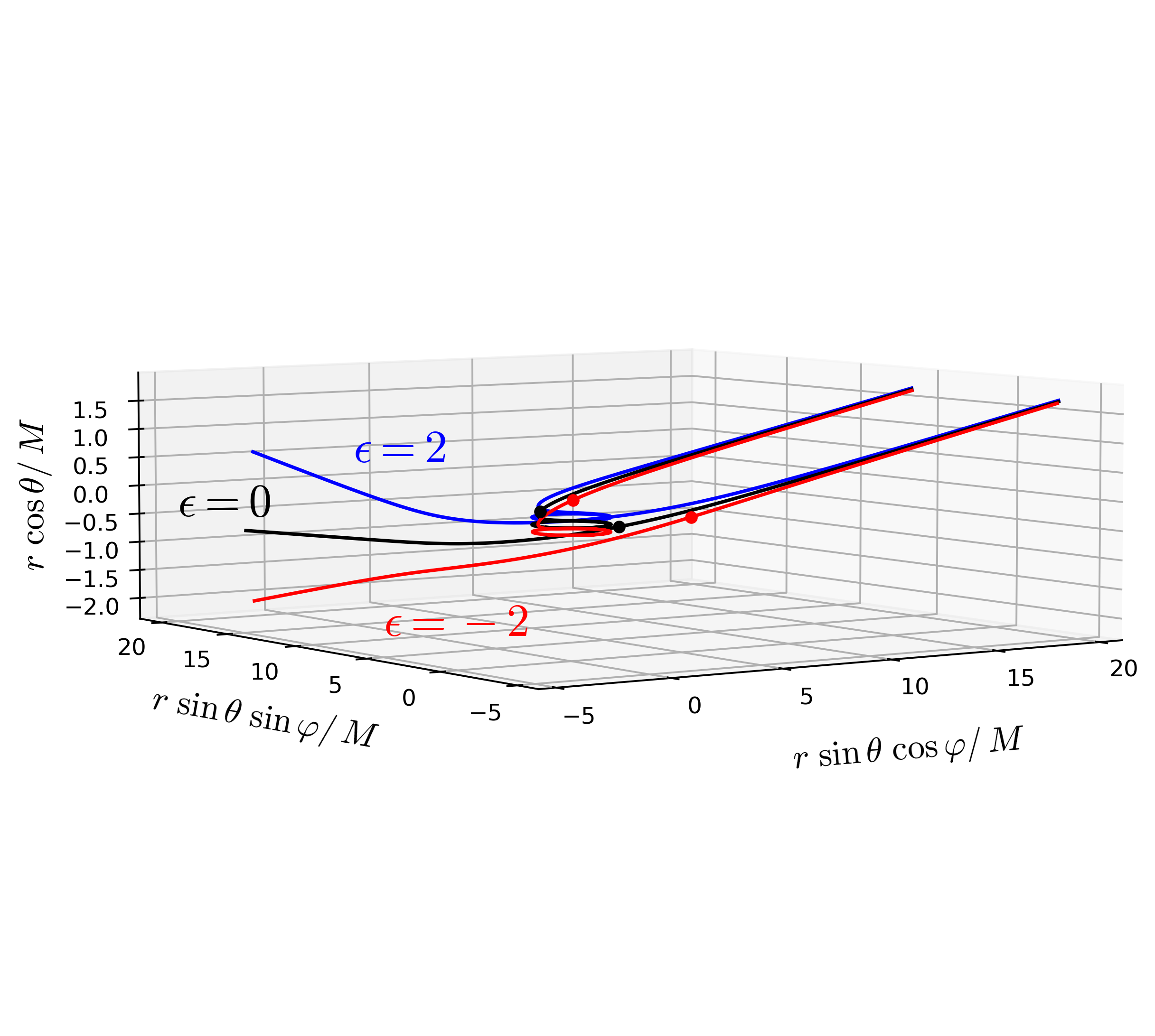}
\caption{Face-on (top panel) and edge-on (bottom panel) views of selected example null geodesics received by a distant observer with $i=85^{\circ}$ in Kerr and No$\mathbb{Z}$ spacetimes, shown in the Boyer-Lindquist coordinates. The observer is located at $\varphi=0$. The geodesics are computed backward in time starting from the observer's image plane, indicated by the `$\times$' and `$+$' symbols shown in Fig.~\ref{fig:disk_z}. The dots represent where the rays are emitted from the Keplerian disk surface (see also Fig.~\ref{fig:disk_z}). For the case of $\epsilon=2$, the geodesics are not emitted from the accretion disk.
}
\label{fig:ray_plane} 
\end{figure} 

\subsection{Keplerian Disk around Black Holes}
Starting from this section, we explore possible observational features of No$\mathbb{Z}$ black hole spacetime beyond the resulting black hole shadow. The reason is twofold. First, we are interested in whether there are striking observational features which may help to tell the No$\mathbb{Z}$ spacetime from Kerr, despite their similar resulting black hole shadow morphologies\footnote{Note that in general a $\mathbb{Z}_2$ asymmetric black hole spacetime may violate integrability symmetry simultaneously, resulting in a very different black hole shadow compared to the cases of Kerr  ~\citep[e.g. Fig. 9(a) of][]{Staelens:2023jgr} and hence easy to tell.} as discussed in sec.~\ref{sec:bh_shadow}. Second, theoretical interpretations of astronomical observations in strong gravity regime should include both the effect of the background spacetime and the features of the luminous materials around the black hole ~\cite[e.g.][]{Kulkarni:2011cy,EventHorizonTelescope:2019pgp,EventHorizonTelescope:2022urf}.

Among the various features of geodesics in the No$\mathbb{Z}$ black hole spacetime compared to the Kerr ones, the fact that the circular orbits around a No$\mathbb{Z}$ black hole are not at the standard equatorial plane is particularly crucial when discussing the disk morphology. For instance, the prograde accretion disk around a Kerr black hole ($\epsilon=0$) is located at the equatorial plane $y=0$ with the ISCO at $\sim 2.32088M$ when $a_*=0.9$. On the other hand, the disk surface around a No$\mathbb{Z}$ black hole with a positive (negative) $\epsilon$ is below (above) the $y=0$ plane (see the upper panel of Fig.~\ref{fig:ryplot}). For the No$\mathbb{Z}$ black hole with $a_*=0.9$ and $\epsilon=\pm2$, the ISCO is located at $\sim2.31789M$. The prograde ISCO radius is shrunk compared with that of the Kerr black hole \cite{Chen:2021ryb}. Since the deviations of ISCO, plane orbits, angular velocity (Fig.~\ref{fig:ryplot}), and the orbital angular momentum (see Appendix \ref{appendix:angular_momentum}) are limited between the Kerr and the No$\mathbb{Z}$ spacetime, we assume astrophysical thin disks around the No$\mathbb{Z}$ black holes under consideration can be formed in a similar way to that around Kerr black holes \citep[e.g.][]{Novikov:1972}.

In this subsection, We consider the No$\mathbb{Z}$ black holes with $a_*=0.9$ and $\epsilon=\pm2$ and $1$. Then we explore the morphology and the local redshifts of a prograde thin disk around the black hole. The thin disk consists of a collection of circular and Keplerian orbits, with the inner edge determined by the ISCO. The energy shift  $E/E_0$ between the frame of distant observers ($r\to\infty$) and the local comoving frame of the emission is introduced by both the four-momentum of photons $p^{\alpha}$ and the disk dynamics, i.e. the four-velocity $u^{\alpha}$ of the Keplerian orbits. More explicitly, we have
\begin{equation}\label{eq:redshift}
    \dfrac{E}{E_{0}}=\dfrac{(p_{\alpha}u^{\alpha})|_{\infty}}{(p_{\alpha}u^{\alpha})|_{\rm 0}}\,,
\end{equation}
where the subscript ``0" denotes the quantities computed in the local comoving frame.

In Fig.~\ref{fig:disk_z}, we present the redshift images and the morphology of a thin disk around a Kerr and No$\mathbb{Z}$ black holes (with $a_*=0.9$) seen by a distant observer observing from four different viewing angles, $i=45^{\circ}, 85^{\circ}, 90^{\circ}, 95^{\circ}$. The accretion disk is assumed to have an infinitesimally small thickness. For all cases, the inner edge of the disk is located at the ISCO and the outer edge of the disk is $r_{\rm out}=10 M$. The color bar indicates the redshifts $E/E_0$. The approaching side of the disk is blue-shifted ($E/E_0>1$) and the receding side is red-shifted ($E/E_0<1$). For reference, the critical curves of shadows are also shown by the green dashed curves.

When $i=90^{\circ}$, the disk image in Fig.~\ref{fig:disk_z} is vertically symmetric for $\epsilon=0$, but not for the No$\mathbb{Z}$ black hole cases. Note also that because the distortion function $\tilde\epsilon(y)$ is invariant under $(\epsilon,y)\rightarrow(-\epsilon,-y)$, the images are vertically reflected under $(\epsilon,i)\rightarrow(-\epsilon,-i)$. Therefore, the disk morphology for $\epsilon=2$ with $i=85^{\circ}, 90^{\circ}, 95^{\circ}$ are simply the vertical flipping of the images for $\epsilon=-2$ with $i=95^{\circ}, 90^{\circ}, 85^{\circ}$, respectively.

In addition, the contrast of the disk images between the Kerr and the No$\mathbb{Z}$ black hole spacetimes becomes more apparent when the accretion disk is observed nearly edge-on. Such a contrast is evident in both the disk morphology and grid distortions. Notably, with $\epsilon=2$ and $i=85^{\circ}$ (or equivalently, with $\epsilon=-2$ and $i=95^{\circ}$), the inner part of the upper disk surface exhibits a visually concave shape on the approaching (left) side. A similar feature is also present for the case of $\epsilon=1$ and $i=85^{\circ}$, but the concave shape is less obvious. This distinctive feature arises from the effects introduced by the $\mathbb{Z}_2$ asymmetry on the geometry of the disk surface, as illustrated in Fig.~\ref{fig:ryplot}, and the corresponding alterations in photon geodesics. 

To elucidate the concave feature for the case of $(\epsilon,~i)=(2,~ 85^{\circ})$ in Fig.~\ref{fig:disk_z}, we pinpoint two rays that correspond to the receptions at $(+4.5, 0)$ and $(-4.5, 0)$ on the image plane, as respectively indicated by the cyan `+' and `$\times$' symbols shown in Fig.~\ref{fig:disk_z}, and show the trajectories of these rays in Fig.~\ref{fig:ray_plane}. 
It is shown that, for both $\epsilon=0$ and $\epsilon=-2$ cases, the rays cross the disk, meaning that observers can receive photons emitted from the disk surface. However, for the case of $\epsilon=2$, the light rays do not cross the disk. Hence, the concave features manifest on the image planes in the scenario where $(\epsilon,~i)=(2,~ 85^{\circ})$ in Fig.~\ref{fig:disk_z}. We note that 
such distinctive features of $\mathbb{Z}_2$ asymmetry in our model are due to the opposite shift between the photon geodesics and circular Keplerian orbits. As we have mentioned at the end of sec.~\ref{sec:th},  the photon geodesics for $\epsilon>0$ are shifted upward, but the Keplerian circular orbits for $\epsilon>0$ are shifted downward with respect to the standard equatorial plane $y=0$ (see also Fig.~\ref{fig:ryplot}).

\begin{figure*}
  \centering
\includegraphics[width=0.95\textwidth]{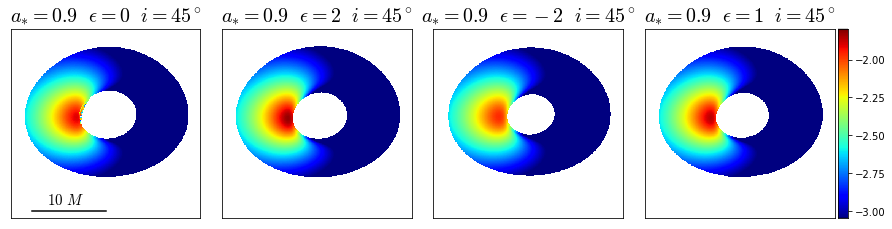}
\includegraphics[width=0.95\textwidth]{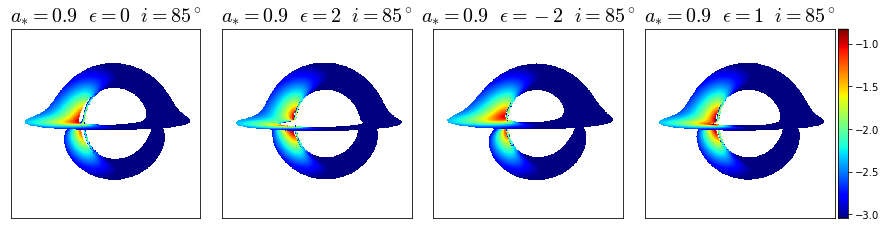}
\includegraphics[width=0.95\textwidth]{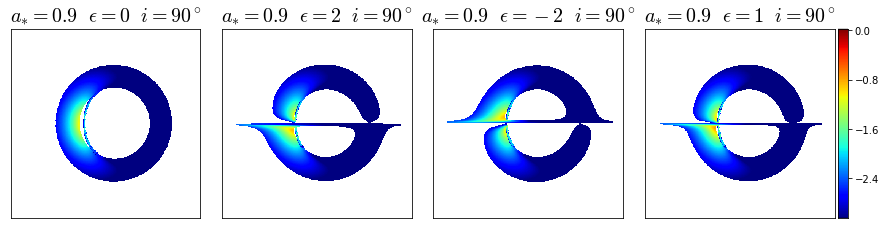}
\includegraphics[width=0.95\textwidth]{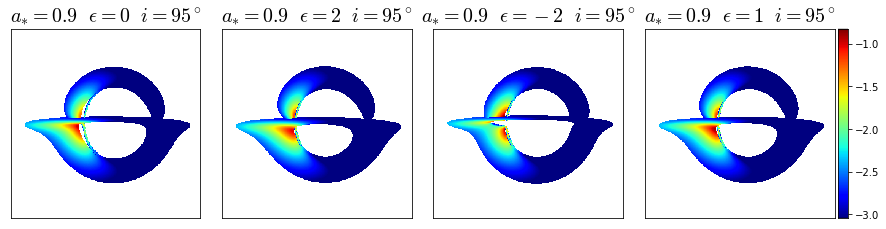}
\caption{The line emission images of the Keplerian disks considered in Fig.~\ref{fig:disk_z}. The colored contour indicates the flux distribution, which is shown in logarithmic scales with relative units. The line emissivity on the disk surface is assumed to be decreasing radially following Eq.~\eqref{eq:power_law} with $q=2$. For comparison, the same normalization applies to images with the same inclination angle, indicated by the color bar shown at the right in each row.}
\label{fig:disk_flux} 
\end{figure*} 
 
\subsection{Line emission and spectral profile features}

Different from directly resolving the images of the accretion disk, observing aggregate emissions emanating from the disk can also be used to probe the relativistic environment of a black hole. For instance, the profile of a relativistically broadened spectral line from the accretion disk of a black hole has been utilized to estimate black hole spins. In this subsection, we are interested in looking for the distinctive observable features in line emission profiles that can be generated by \ztwo asymmetry.

To compare the line profile features in Kerr and  No$\mathbb{Z}$ black hole spacetimes, we assume that the local line emissivity, $j_{0}$,  on the surface of an optically thick accretion disk decreases radially outward in a power-law manner, i.e., 
\begin{equation}\label{eq:power_law}
j_{0}\propto r^{-q}\,.
\end{equation}
The disk line profiles and the associated flux images are computed by considering the line emission emitted from the disk plane, taking into account the effects of photon geodesics, energy shift, and disk emissivity all together with general relativistic radiative transfer (e.g., \cite{Fuerst:2004ii}). 

Assuming $q=2$, in Fig.~\ref{fig:disk_flux} we present the line emission images with the same parameters as those in Fig.~\ref{fig:disk_z}. The color bar indicates the relative flux with the same normalization for all the images in a row. It is evident that the emission is mostly contributed from the blue-shifted side of the disk, and the blue-shifted (${E}/E_{0}>1$) flux is enhanced as the disk approaches edge-on orientations. The disparities in flux distribution between the No$\mathbb{Z}$ black hole spacetime and the Kerr one become particularly pronounced in such orientations.

\begin{figure*}
  \centering
\includegraphics[width=0.45\textwidth]{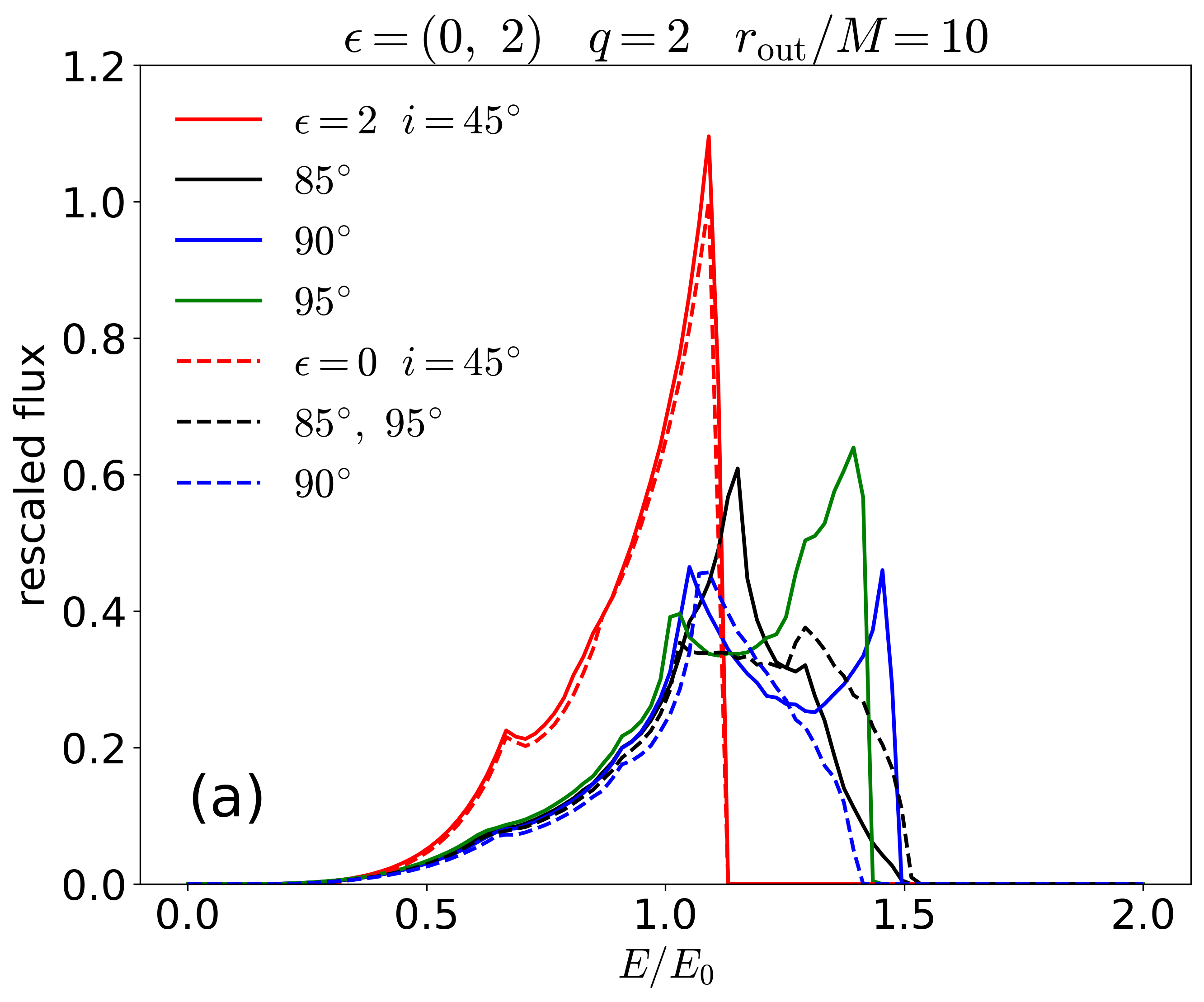}
\includegraphics[width=0.45\textwidth]{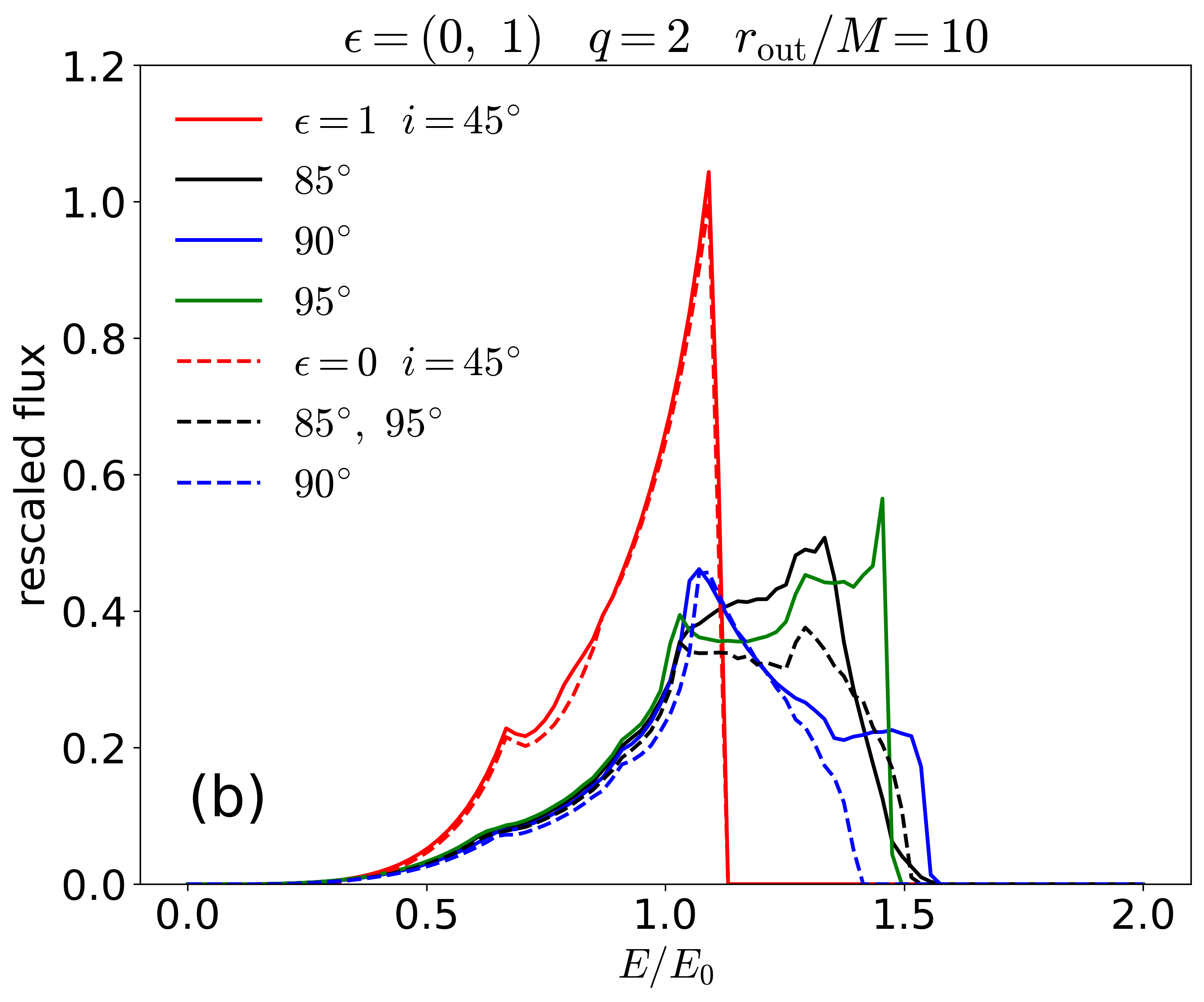}\\
\includegraphics[width=0.45\textwidth]{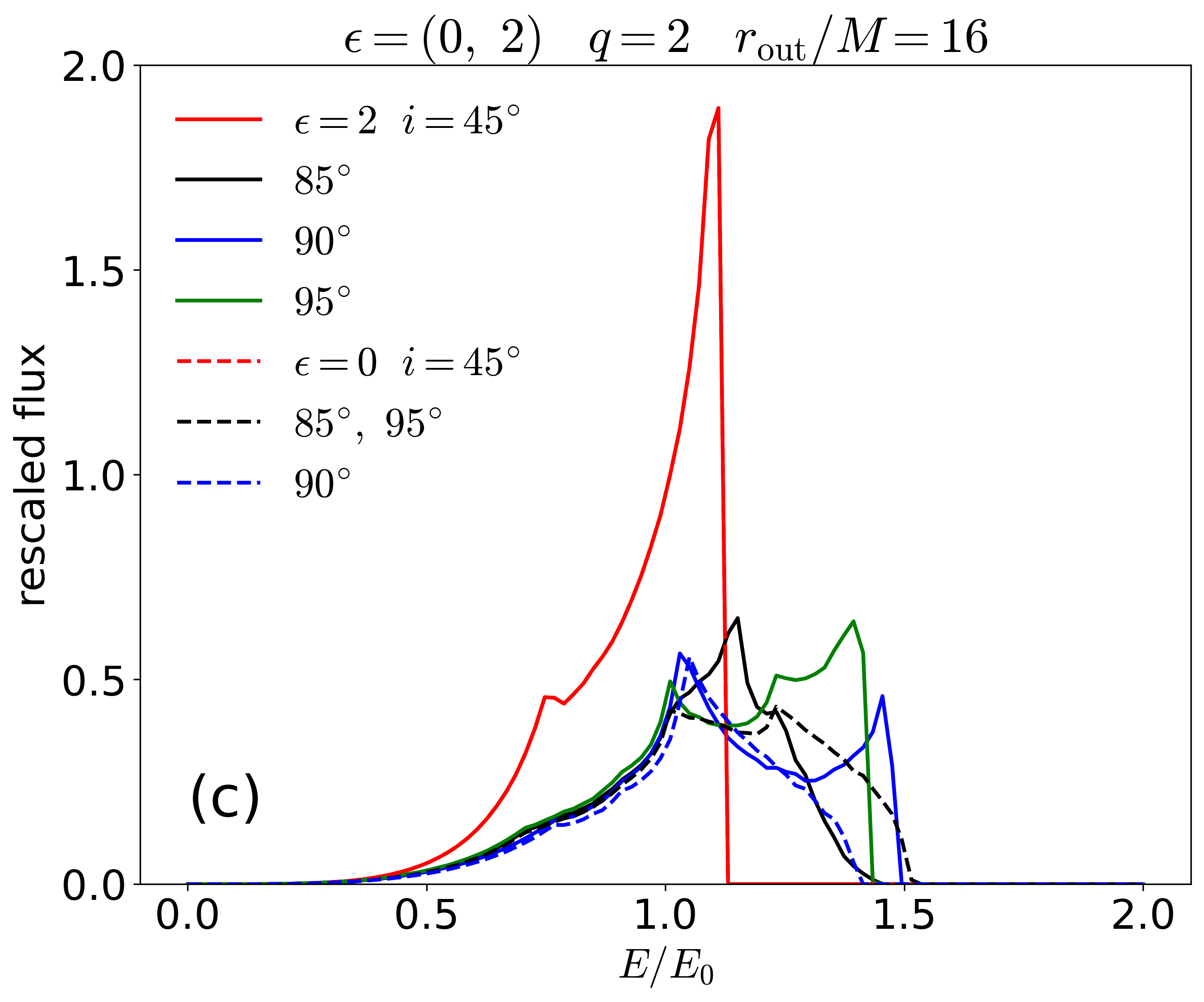}
\includegraphics[width=0.45\textwidth]{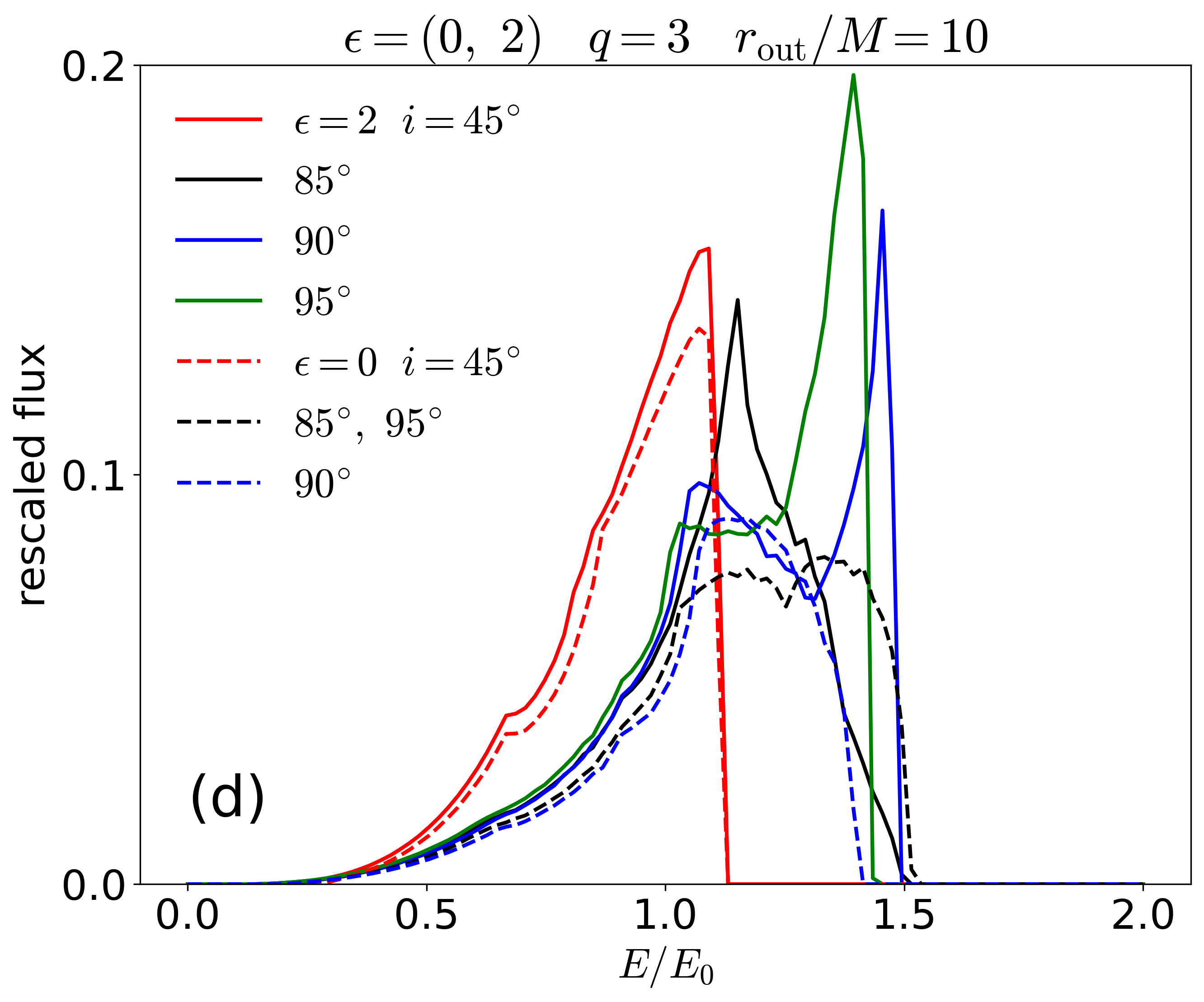}\\
\caption{Line profiles for $a_{*}=0.9$ and different $(\epsilon,~q,~r_{\rm out},~i)$. For comparison, the line profiles of the Kerr case ($\epsilon=0$) are shown by the dashed lines.  All the line profiles are normalized to the peak flux of the line with  $(\epsilon,~q,~r_{\rm out},~i)$=$(0,~2,~10M,~45^{\circ})$.}
\label{fig:line_profile}
\end{figure*}

Combining the information of the redshift images in Fig.~\ref{fig:disk_z} and the line emission images in Fig.~\ref{fig:disk_flux}, we can compute the spectral line profiles. The modeled line profiles can be computed by integrating the flux of each pixel of the image at different energy shifts, with $100$ bins between the range $0<({E}/E_{0})<2$.  We validate our computations by reproducing the results in Ref.~\cite{Fuerst:2004ii}. In Fig.~\ref{fig:line_profile}, we adopt $a_{*}=0.9$ and compare the line profiles when different parameters, $\epsilon$, $q$, $r_{\rm out}$, and the inclination angle $i$, are considered. To indicate the relative strength of the line profiles, all the line profiles are normalized to the peak flux of the line with  $(\epsilon,~q,~r_{\rm out},~i)$=$(0,~2,~10M,~45^{\circ})$.

The line profiles in the cases of $\epsilon=2$ and $\epsilon=0$ (as shown in the first two columns of Fig.~\ref{fig:disk_flux}) are shown in  Fig.~\ref{fig:line_profile}(a). We compare the results and enumerate some interpretations  as follows:
\begin{enumerate}
\item Due to the \ztwo symmetry of the Kerr spacetime, the profiles for $(\epsilon,~i) =(0,~85^{\circ}/95^{\circ})$ are the same (the black dashed lines). However, this is not the case when \ztwo symmetry is broken, e.g. the profiles for $(\epsilon,~i) =(2,~85^{\circ}/95^{\circ})$ are different, as one can see from the solid black and green lines. 
\item The discrepancies between $\epsilon= (0,~2)$ are mostly obvious at the blue-shifted part of the profiles ($E/E_{0}>1$). This is because most of the flux in an image (Fig.~\ref{fig:disk_flux}) is contributed from the blue-shifted part (Fig.~\ref{fig:disk_z}). In comparison, the profiles are alike at the red-shifted part ($E/E_{0}<1$). This implies that the feature of \ztwo asymmetry mainly comes from the approaching side of the inner disk, which is consistent with what we observed from Figs.~\ref{fig:disk_z} and \ref{fig:disk_flux}.
\item Comparing the red lines and those with other colors, one can see that the discrepancies between $\epsilon= (0,~2)$ are mostly obvious at higher inclination angles, namely, when the inclination is almost edge-on.
\item Both the line profiles of $(\epsilon,~i) =(0,~90^{\circ})$ and $(\epsilon,~i) =(2,~90^{\circ})$ has a horn at $E/E_{0}\gtrsim 1$. However, in the latter case, there is an additional flux contribution at a relatively higher value of $E/E_{0}$ (the third row in Fig.~\ref{fig:disk_flux}), resulting in another horn in the line profiles near $E/E_{0}\sim 1.5$. Similarly, the line profiles for $(\epsilon,~i) =(0,~95^{\circ})$ and $(\epsilon,~i) =(2,~95^{\circ})$ are similar up to $E/E_{0}\gtrsim 1$, and the latter shows an additional horn feature at the higher end to $E/E_{0}$.

\item Comparing the line profiles between $(\epsilon,~i) =(0,~85^{\circ})$ and $(\epsilon,~i) =(2,~85^{\circ})$, an obvious horn feature shows up at $E/E_{0}\gtrsim 1$ in the latter case. This horn feature is different from those appearing at $E/E_{0}\sim 1.5$ in the cases of $i=90^{\circ}$ or $i=95^{\circ}$. In particular, near $E/E_{0}\sim 1.5$, the flux of the line profile with $(\epsilon,~i) =(2,~85^{\circ})$ is lower than that of $(\epsilon,~i) =(0,~85^{\circ})$,
due to that the appearance of the concave shape in the image (Fig.~\ref{fig:disk_flux}) reduces the flux contributed by the higher end of $E/E_0$ for the No$\mathbb{Z}$ case.
\end{enumerate}

In Fig.~\ref{fig:line_profile}(b), we present the line profiles for the cases of $\epsilon =1$ and $\epsilon =0$ (see also the first and the last columns of Fig.~\ref{fig:disk_flux}). As the \ztwo asymmetry becomes less obvious when $|\epsilon|$ decreases,  the differences between the horn flux resulting from the \ztwo asymmetry and the flux for $\epsilon=0$ case are less significant, as compared with the results in Fig.~\ref{fig:line_profile}(a). 

Finally, we check how unique the novel horn features in the line profiles are associated with the \ztwo asymmetry. In Figs.~\ref{fig:line_profile}(c) and (d), we keep all parameters the same as those in Fig.~\ref{fig:line_profile}(a), except for one test parameter. Fig.~\ref{fig:line_profile}(c) displays the case when the disk has a larger outer edge, $r_{\rm out}=16M$. In this case, the flux in the line profile at $i=45^{\circ}$ is significantly larger than that of $r_{\rm out}=10M$ (Fig.~\ref{fig:line_profile}(a)), due to that a much larger disk contributes more flux at $E/E_{0}\lesssim1$. On the contrary,  the line profiles at more edge-on angles, $i=(85^{\circ},~90^{\circ},~95^{\circ})$, remain quantitatively similar. Fig.~\ref{fig:line_profile}(d) shows the case when the emissivity decays with radius more rapidly, i.e., $q=3$.  Compared with the $q=2$ case (Fig.~\ref{fig:line_profile}(a)), the line flux is weaker but concentrates more at a smaller radius, where the emission of higher $E/E_{0}$ is radiated. As a result, the differences between the horn flux resulting from the \ztwo asymmetry and the $\epsilon=0$ case are enhanced further. We conclude that the horn features induced by the \ztwo asymmetry are quite distinctive in the sense that they are not sensitive to the changes of $q$ and $r_\textrm{out}$.

\section{\label{sec:con}Conclusions}
As one of the fundamental symmetries of the Kerr spacetime, the violation of \ztwo asymmetry of an isolated black hole directly implies the violation of the Kerr hypothesis in GR. In this work, we explore the possible observational features of \ztwo asymmetric black hole spacetimes. In particular, we focus on the No$\mathbb{Z}$ black hole spacetime proposed in Refs.~\cite{Chen:2020aix,Chen:2021ryb,Chen:2022lct}, and investigate the morphology and the line profiles of black hole accretion disks by employing numerical computations to trace photon geodesics and radiative transfer of light rays emitted from the disk. 

Within this No$\mathbb{Z}$ \ spacetime framework,  we observe a distinct shift in the trajectories of photon geodesics, and circular Keplerian orbits towards different directions with respect to the standard equatorial plane  $y=0$. Consequently, a concave distortion emerges at the approaching side of the inner edge on the disk image, particularly noticeable when the inclination angle $i$ is approximately less than $90^{\circ}$ for $\epsilon>0$, and greater than $90^{\circ}$ for $\epsilon<0$. Such features of \ztwo asymmetry become more pronounced with an increasing black hole spin and higher values of $|\epsilon|$.

By assuming that the line emission from the disk surface is decreasing radially outward, we further find that the effect induced by  \ztwo asymmetry is mainly on the blue-shifted ($E/E_0>1$) part of the line profiles. The discrepancy between the line profiles in the Kerr and the No$\mathbb{Z}$ black hole spacetimes is thus most significant when the inclination angle is almost edge-on, i.e., $i\sim90^{\circ}$, manifesting as distinct horn-like features at the blue-shifted extremity of the line profile for the No$\mathbb{Z}$ spacetime. Remarkably, the emergence of a concave shape in the inner region of the approaching side on the disk images leads to a migration of these blue horns in the line profile toward $E/E_0\rightarrow1$. This shift occurs because the blue-shifted line emission diminishes relatively due to the presence of the concave shape at the inner edge of the disk image. These horn-like features in the line emission are not sensitive to the changes in the emission profile and the disk size, as demonstrated in Figs.~\ref{fig:line_profile}(c) and (d), suggesting that these features are quite specific to the \ztwo asymmetry of the spacetime. Our findings also present an intriguing illustration of discernible observable features that can be used to distinguish between different spacetime models, even when the size and the shape of black hole shadows remain similar among these spacetime configurations \cite{Vagnozzi:2022moj,Ayzenberg:2023hfw,Cardenas-Avendano:2023obg}.

We would like to emphasize that in order to have noticeable $\mathbb{Z}_2$ asymmetric features in the line profiles mentioned above, the inclination angle has to be nearly edge-on. Also, the spin value of the black hole and the values of $|\epsilon|$ have to be sufficiently large in order to have significant features. In particular, the concave shape in the images of the disk morphology requires that the Keplerian disk and the light rays are shifted in opposite directions, which is certainly a model-dependent feature. It remains to be explored in the future whether there are other unique features of $\mathbb{Z}_2$ asymmetry that are more model-independent and may even be observable in lower inclination angles or lower spins.

\begin{appendix}

\section{The construction of the No$\mathbb{Z}$ black hole metric}\label{appeid.a}

In this appendix, we review the construction of the phenomenological model of the No$\mathbb{Z}$ black hole spacetime, whose metric is given by Eqs.~\eqref{gtt}-\eqref{g23}. The model was proposed in Refs.~\cite{Chen:2020aix,Chen:2021ryb} with the assumption that, among the symmetries owned by the Kerr spacetime, only the equatorial reflection symmetry is allowed to be broken. Therefore, the resulting No$\mathbb{Z}$ spacetime allows us to look for the observational features that are purely induced by the violation of $\mathbb{Z}_2$ symmetry. In addition, we also require the resulting spacetime to be asymptotically flat.

In order to preserve the hidden symmetry that corresponds to the separability of geodesic equations, we consider the general axisymmetric and stationary metric representation proposed in Ref.~\cite{Papadopoulos:2018nvd} (see also Ref.~\cite{Benenti:1979erw}). The metric in its contravariant form, in the Boyer-Lindquist coordinates $(t,r,y,\varphi)$ can be expressed as follows
\begin{align}
g^{tt}&=\frac{\mathcal{A}_5(r)+\mathcal{B}_5(y)}{\mathcal{A}_1(r)+\mathcal{B}_1(y)}\,,\qquad g^{t\varphi}=\frac{\mathcal{A}_4(r)+\mathcal{B}_4(y)}{\mathcal{A}_1(r)+\mathcal{B}_1(y)}\,,\nonumber\\
g^{\varphi\varphi}&=\frac{\mathcal{A}_3(r)+\mathcal{B}_3(y)}{\mathcal{A}_1(r)+\mathcal{B}_1(y)}\,,\qquad g^{yy}=\frac{\mathcal{B}_2(y)}{\mathcal{A}_1(r)+\mathcal{B}_1(y)}\,,\nonumber\\
g^{rr}&=\frac{\mathcal{A}_2(r)}{\mathcal{A}_1(r)+\mathcal{B}_1(y)}\,,\label{PKmetric}
\end{align}
where $A_i(r)$ and $B_i(y)$ are functions of $r$ and $y$, respectively. As long as the spacetime metric can be recast in the form of Eq.~\eqref{PKmetric}, there exists a non-trivial rank-2 Killing tensor, and therefore the geodesic dynamics is Liouville integrable. 

Relaxing the $\mathbb{Z}_2$ symmetry only requires the metric functions to contain some terms that are not symmetric with respect to $y=0$. For the metric to be a ``minimal modification" of the Kerr spacetime, we assume that the radial metric functions are exactly given by their Kerr counterparts:
\begin{align}
\mathcal{A}_1=r^2\,,\quad \mathcal{A}_2=\Delta\,,\quad \mathcal{A}_3=-\frac{a^2}{\Delta}\,,\nonumber\\
\mathcal{A}_4=-\frac{aX}{\Delta}\,,\qquad\mathcal{A}_5=-\frac{X^2}{\Delta}\,,\label{Kerrianmetric}
\end{align} 
where $\Delta\equiv r^2-2Mr+a^2$ and $X\equiv r^2+a^2$. On the other hand, we assume that the polar metric functions $\mathcal{B}_i(y)$, on top of their Kerr counterparts, contain some deviation functions \cite{Chen:2020aix}:
\begin{align}
&\mathcal{B}_1=a^2y^2+\tilde\epsilon_1(y)\,,\qquad \mathcal{B}_2=1-y^2\,,\qquad\mathcal{B}_3=\frac{1}{1-y^2}\,,\nonumber\\
&\mathcal{B}_4=a\,,\qquad \mathcal{B}_5=a^2(1-y^2)+\tilde\epsilon_5(y)\,,
\end{align}
where we have assumed that the spacetime is asymptotically flat such that the deviation functions in $\mathcal{B}_3(y)$ and $\mathcal{B}_4(y)$ vanish. Also, the deviation function in $\mathcal{B}_2(y)$ can be removed through a coordinate transformation. Therefore, only $\tilde{\epsilon}_1(y)$ and $\tilde{\epsilon}_5(y)$ remain. 

Furthermore, we may have one additional constraint from the Solar System test. The observational requirements that the post-Newtonian parameters $\beta$ and $\gamma$ should be very close to one give \cite{Chen:2021ryb,Chen:2022lct}
\begin{equation}
\left|\tilde\epsilon_1(y)+\tilde\epsilon_5(y)\right|\ll M^2\,.
\end{equation} 
Therefore, we shall assume that $\tilde\epsilon_1(y)=-\tilde\epsilon_5(y)=\tilde\epsilon(y)$. As long as the deviation function $\tilde\epsilon(y)$ is well-defined on the axis of symmetry, i.e., $y=\pm1$, it can be shown that 
\begin{equation}
\frac{\mathcal{X}_{,\mu}\mathcal{X}^{,\mu}}{4\mathcal{X}}\Bigg|_{y\rightarrow\pm1}\rightarrow1\,,
\end{equation}
where $\mathcal{X}\equiv\eta_\mu\eta^\mu=g_{\varphi\varphi}$ is the norm of the Killing vector $\eta=\partial_\varphi$. According to Ref.~\cite{Stephani:2003tm}, the spacetime has no conical singularity on the axis of rotation even if the $\mathbb{Z}_2$ symmetry is broken.{\footnote{We thank David Kofro\v{n} for pointing out this issue.}}

Finally, to accommodate the features that the $\mathbb{Z}_2$ asymmetry of black hole models is usually associated with a nonzero black hole spin in parity-violating theories \cite{Cardoso:2018ptl,Cano:2019ore,Cano:2022wwo,Tahara:2023pyg}, we assume that the deviation function $\tilde{\epsilon}(y)$ is proportional to the black hole spin $a$, and has a linear dependence on $y$:
\begin{equation}
\tilde{\epsilon}(y)=\epsilon May\,,
\end{equation}
where $\epsilon$ is a dimensionless constant. This choice of the deviation function can be regarded as the first-order correction from some putative parity-violating effects from a phenomenological point of view. 

\section{Specific angular momentum of thin disk in the No$\mathbb{Z}$ black hole metric}\label{appendix:angular_momentum}
\begin{figure}[t]
  \centering
 \includegraphics[scale=0.6]{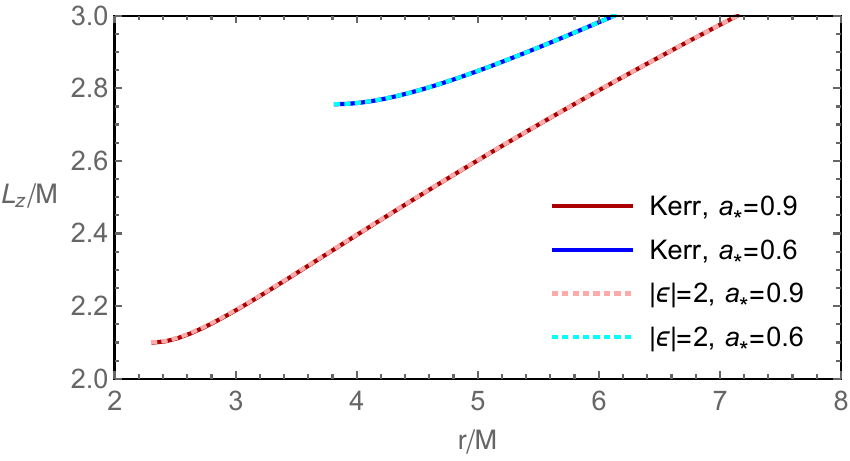}
\\
\caption{The specific angular momentum $L_z$ of prograde circular orbits as a function of radial coordinate $r$. The profiles end at the ISCO.}
\label{fig:lzplot} 
\end{figure}

In Fig.~\ref{fig:lzplot}, we compare the specific angular momentum profiles, $L_{z}/M\equiv u^{\phi}$, of a thin disk in Kerr and the No$\mathbb{Z}$ spacetime, with the same parameters shown in the bottom panel of Fig.~\ref{fig:ryplot}. The ISCOs are located at the minimal value of the profiles, which are almost overlapped for the Kerr and No$\mathbb{Z}$ cases.

\end{appendix}

\section*{Acknowledgement}
CYC is supported by the Special Postdoctoral Researcher (SPDR) Program at RIKEN. HYP is supported by 
Yushan Young Fellow Program, Ministry of Education (MOE), Taiwan, and the National Science and Technology Council (NSTC), Taiwan, under the grant 112-2112-M-003-010-MY3.

\end{document}